\documentclass[journal abbreviation, manuscript]{copernicus}

\begin{document}
\nolinenumbers

\title{
GPROF V08: Improving Satellite Precipitation Estimates Across a Heterogeneous Constellation of Passive-Microwave Sensors
}

\Author[1][simon.pfreundschuh@colostate.edu]{Simon}{Pfreundschuh} 
\Author[1]{Paula J.}{Brown}
\Author[1]{Christian D.}{Kummerow}

\affil[1]{Department of Atmospheric Science, Colorado State University, Fort Collins, CO 80523, USA}




\runningtitle{GPROF V08}

\runningauthor{Simon Pfreundschuh}

\received{}
\pubdiscuss{} 
\revised{}
\accepted{}
\published{}


\firstpage{1}

\maketitle

\begin{abstract}

Precipitation estimates from the passive microwave observations (PMW) of the
Global Precipitation Measurement (GPM) mission serve a wide range of
scientific, societal, and economic applications. The Goddard Profiling Algorithm
(GPROF) is the operational precipitation retrieval used to produce the level-two
precipitation estimates for the GPM mission. This study presents GPROF V08 --
the most recent version of GPROF and a major update aimed at improving the
accuracy of precipitation estimates from the GPM constellation.

GPROF V08 replaces the single-pixel Bayesian retrieval scheme used by earlier
GPROF versions with a convolutional neural network (CNN). Implementing the
algorithm for the full GPM constellation required addressing several challenges
such as the generation of synthetic observations suitable for training CNN
retrievals, which has been described in the first part of this two-part study.
This second part describes the final, operational implementation of GPROF V08,
which adds a finetuning stage to the training to overcome remaining
uncertainties in the synthetic training data. We extensively validate the new
GPROF V08 algorithm for principal GPM sensors against reference precipitation
estimates from combined radar and radiometer retrievals from GPM and TRMM as
well as independent ground-based reference measurements. While the magnitude of
improvements varies by sensor and surface type, the retrieval accuracy for all assessed
sensors is improved.

Although  challenges such as the persistence of climatological biases
(which remain of the same order as for GPROF V07) persist, GPROF V08 constitutes a
substantial advancement in spaceborne precipitation monitoring. Since the
improvements are robust across all assessed GPM sensors, the update is expected
to improve the full historical GPM precipitation record and thus benefit a wide
range of downstream datasets and applications.

\end{abstract}

\introduction  

The Global Precipitation Measurement (GPM, \citealp{hou_global_2014}) mission
led by NASA and JAXA is the central international effort to monitor
precipitation globally at hourly timescales. To achieve this, the mission
leverages observations from a constellation of passive microwave sensors. The
Goddard Profiling Algorithm (GPROF, \citealp{kummerow_evolution_2015,
randel2020}) is the operational precipitation retrieval algorithm used to
produce precipitation estimates from the PMW observations of all constellation
sensors. The resulting level-two precipitation estimates are used directly but
also serve as input streams to higher-level datasets such as Integrated
Multi-Satellite Retrievals for GPM (IMERG, \citealp{huffman_integrated_2020})
and the Global Precipitation Climatology Project (GPCP,
\citealp{huffman_new_2023}). The GPROF algorithm is central to the GPM
precipitation products and thus indirectly supports various downstream
applications ranging from meteorology and climate science over flood monitoring
\citep{kirschbaum_2017} to risk and resource management \citep{portier_2023}.

Due to the fundamental role of GPROF within the global precipitation observing
system, there has been continuous development aimed at improving the algorithm
\citep{kummerow_evolution_2015, randel2020}. This paper presents GPROF V08 --
the latest step in the evolution of the GPROF retrieval algorithm. GPROF V08
represents a fundamental change from previous GPROF versions by replacing the
traditional Bayesian retrieval framework \citep{kummerow_evolution_2015} with a
machine-learning retrieval based on a convolutional neural network (CNN). The
transition from the pixel-based Bayesian scheme to a spatially-aware CNN
constitutes a fundamental shift in retrieval paradigm. The application of a CNN
allows the retrieval model to learn leverage the spatial features in the input
observations, allowing the model to provide more accurate precipitation
estimates than pixel-based methods \citep{pfreundschuh_gprof-nn_2022,
pfreundschuh_gprof_2024}.

However, the adoption of CNN-based retrievals to the full GPM constellation,
which at the time of this writing comprises twelve different sensor types across
31 satellites, also presented unique challenges. One of them was the development
of a new simulation framework for generating synthetic brightness temperatures.
Synthetic observations for each sensor are needed to construct large,
climatologically consistent training datasets based on a common precipitation
reference provided by the combined radar–radiometer retrievals from the GPM Core
Observatory. The development and evaluation of this simulation framework are
described in the first part \citep{pfreundschuh_gprofv08_part1} of this two
part study.

While the new simulation framework was found to be sufficiently accurate to
provide improved precipitation retrievals across most constellation sensors,
residual uncertainties in the synthetic observations were found to lead to
retrieval biases for certain sensors. Additionally, it was found that the novel
machine-learning-based retrieval showed larger sensitivity to sensor anomalies
such a noise degradation or channel loss. To address these challenges, the
operational GPROF V08 retrieval further introduces an additional finetuning
stage. In this stage, a retrieval model previously pre-trained on synthetic
observations is finetuned using a smaller dataset of collocations between the
target sensor and GPROF V08 retrievals from a reference sensor.

This study describes the final operational implementation of GPROF V08, which
will be used to produce the level-two PMW precipitation estimates for the next
version of the GPM precipitation products. The retrievals for GMI and the
principal GPM constellation sensors operational prior to 2024 are evaluated
against reference precipitation estimates from various sources in order to
validate the GPROF V08 retrieval and quantify the improvements relative to GPROF
V07.

\section{Methods and Data}

As of 2026, the operational GPROF V08 retrieval produces precipitation estimates
for 11 sensor types flown aboard 31 satellites. Because sensor properties vary
strongly across the sensors, the GPROF V08 retrieval uses a separate model for
each sensor type. While each model is trained using a sensor-specific training
dataset, the generation of training datasets, the model architecture, as well as
the training follows a shared, parametrized processing and training pipeline.
Below, we describe this training pipeline in detail that is applied to train the
GPROF V08 retrieval models for all sensors of the GPM constellation.

\subsection{Neural Network Model}

\begin{figure}
  \centering
  \includegraphics[width=\textwidth]{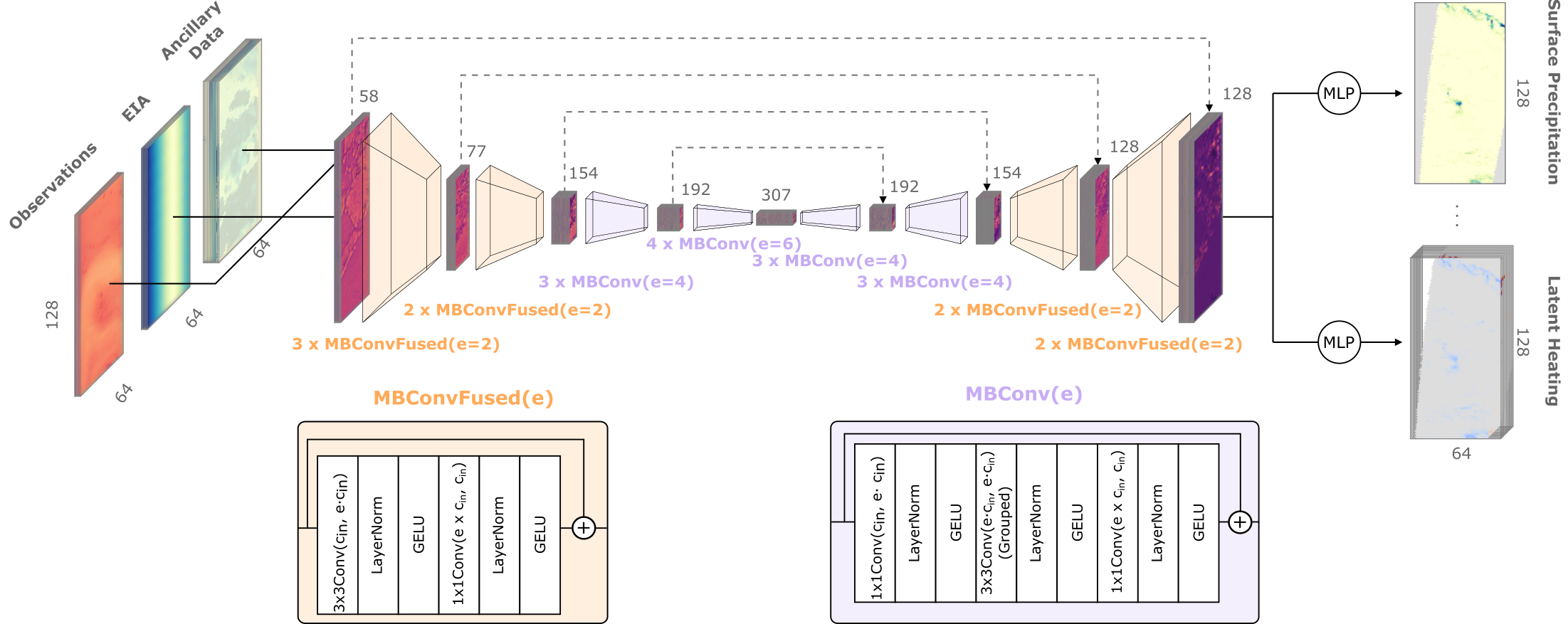}
  \caption{
  The updated neural-network architecture of the GPROF-NN 3D retrieval. The model
  input consists of brightness temperature from 15 channels, corresponding
  earth-incidence angles, and up to 13 different ancillary variables. An U-Net-type
  encoder–decoder model maps these inputs to a shared latent representation from
  which separate multi-layer perceptrons produce estimates of the surface
  precipitation and hydrometeor profiles.
  }
  \label{fig:architecture}
\end{figure}

The GPROF-NN 3D CNN model has been updated slightly compared to the first
version introduced in \citet{pfreundschuh_gprof-nn_2022}. The new architecture
uses inverted-bottleneck convolutional blocks inspired by the MBConv blocks from
the EfficientNet-V2 architecture \citep{tan_efficient_net_2022}. The MBConv
blocks are used without squeeze-and-excite blocks and use GELU activation
functions and layer normalization as proposed in \citet{liu2022convnet}. The
architecture was chosen since it achieved higher accuracy at lower parameter
counts than a comparable standard U-Net model. The input data consists of up to
brightness temperatures from 15 standardized channel slots, corresponding
earth-incidence angles, and ancillary data from up to 14 different variables.
The input tensors are concatenated and passed through the shared U-Net backbone.
Estimates of the observed surface precipitation and hydrometeor profiles are
derived from the shared latent state using separate, pixel-wise fully-connected
heads. The model architecture is illustrated in Fig.~\ref{fig:architecture}.

The retrieval produces quantitative estimates of surface precipitation,
convective precipitation, rain water path, ice water path, cloud water path,
rain water content, snow water content, cloud water content, and latent heating
rates. All profile estimates are produced at 28 vertical levels relative to the
surface with a resolution of 500 m up to 10 km and 1 km above. All quantitative
output variables are predicted using 32 quantiles of the posterior distribution
following \citet{pfreundschuh_neural_2018}. The predicted quantiles cover the
quantile fractions $\tau_i = \frac{i}{33}$ for $i = 1,\ldots, 32$. To
approximate the CDF, which is required to calculate the posterior mean, the CDF
is extended out to 0 and 1 by assuming the CDF gradient outside the last
predicted quantiles is half of the gradient between the last and second-to-last
quantiles. For surface precipitation, the retrieval produces the probability of
precipitation and the first and second tercile, which are derived from the
predicted posterior cumulative distribution function. For all other quantities
only the posterior mean is reported.

\subsection{The GPROF V08 Training Pipeline}

\begin{figure}
  \centering
  \includegraphics[width=\textwidth]{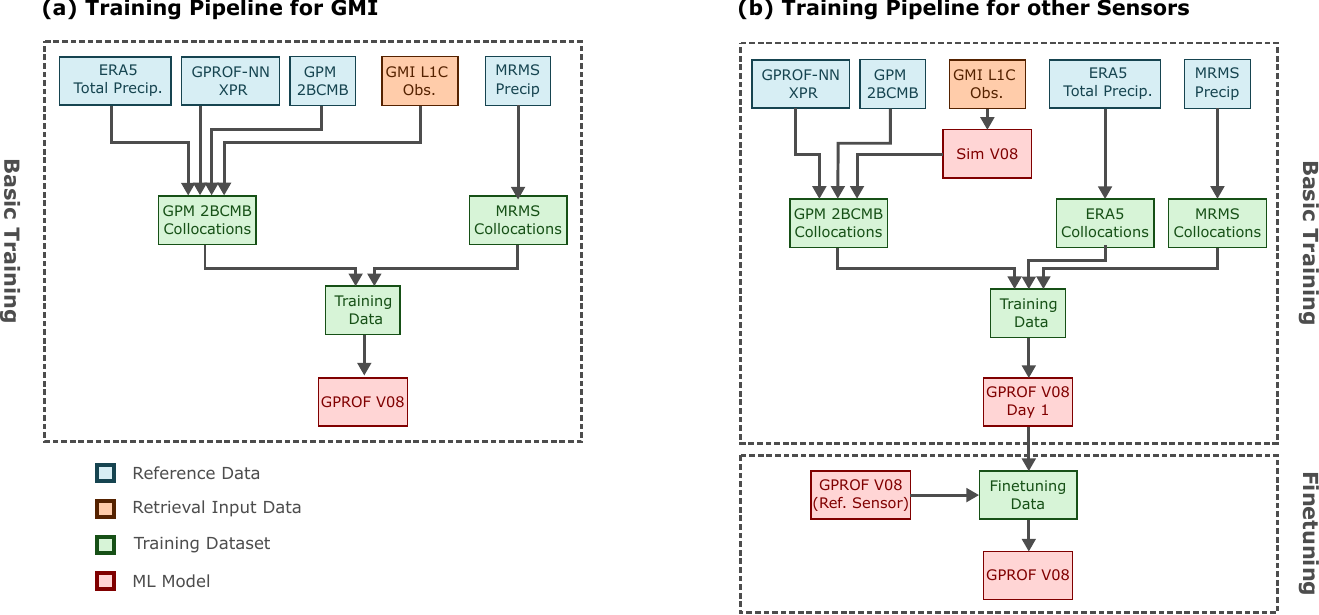}
  \caption{
    Overview of the GPROF V08 training pipeline. For GMI, the retrieval is trained using collocations with the GPM combined radar and radiometer product and ground-based radar measurements from the NOAA's Multi-Radar Multi-Sensor System (MRMS) product over snow-covered surfaces. For other constellation sensors, the same reference precipitation estimates from the combined radar-radiometer retrievals are used with synthetic observations from the Sim V08 simulator. Training on the basic training data produces the GPROF V08 Day-1 retrieval for the target sensor. For sensors that overlap with GMI or another suitable reference sensor, an additional finetuning step is applied that uses actual instead of simulated observations.
  }
  \label{fig:training_strategy}
\end{figure}

 An overview of the training pipeline is shown in Fig.~\ref{fig:training_strategy}.
For all sensors except GMI and AMSR-E, the training of the GPROF V08 retrievals
is performed in two stages: A basic stage and a finetuning stage.

\subsubsection{Basic Training}

The main part of the training data for the basic training is derived from
precipitation estimates from the GPM combined radar and radiomenter retrieval
(2BCMB; \citealp{grecu_gpm_2016}). The 2BCMB data is complemented with
precipitation estimates from additional sources to address known shortcomings of
the 2BCMB precipitation estimates. In general, the basic training data for each
sensor is composed of three subsets:

\begin{description}
    \item[GPM 2BCMB Collocations] This dataset consists of coincident
observations from the GMI sensor matched with precipitation estimates from the
GPM 2BCMB product from a database period that is the same for all sensors.
Similar to GPROF V07, GPROF V08 uses October 2018 to September 2019 as the
database period. Over ocean surfaces, the 2BCMB reference data are augmented by adding
light precipitation from the GPROF-NN XPR retrieval \citep{gprof_nn_xpr} to
improve the sensitivity and climatological representation of light rain and
snow. Additionally, precipitation from ERA5 \citep{hersbach_era5_2020} is added
in very dry and cold environments (two-meter temperature below $260\ \unit{K}$
and total-column water vapor less than $6\ \unit{kg\ m^{-2}}$) when ERA5
suggests precipitation but the 2BCMB is not. Over mountainous regions, the 2BCMB
reference precipitation is enhanced to match accumulations from ERA5 to account
for orographic effects not captured by the reference data. The training
data for non-GMI sensors is produced using the same GMI-2BCMB collocations but
uses the Sim V08 simulator described in \citet{pfreundschuh_gprofv08_part1} to
produce synthetic observations for the target sensor.
  \item[MRMS Collocations] This dataset consists of coincident observations from
  the target sensor and gauge-corrected ground-based radar measurements from
  NOAA’s Multi-Radar Multi-Sensor System (MRMS, \citealp{smith_multi-radar_2016}). These collocations are only extracted
  over snow-covered surfaces and only pixels that are indicated to be snow-covered
  are used for training. Non-GMI sensors use sensor specific collocation datasets.
  For sensors for which no MRMS data is available, observations are adapted from a
  surrogate sensor with similar observation characteristics.
  \item[ERA5 Collocations] Due to the lack of reliable reference precipitation
estimates over sea ice, GPROF uses ERA5 precipitation estimates as training
targets over sea ice, Greenland, and Antarctica. ERA5 collocations are
observations from the target sensor matched with ERA5 precipitation. These
collocations are extracted only over sea ice and only pixels covered by sea ice
are used during training.
\end{description}

The training approach differs slightly between GMI and other sensors of the GPM
constellation. Since GMI is permanently collocated with the combined radar and
radiometer estimates from the 2BCMB product, it can be trained using actual
observations and does not require the use of synthetic observations from the Sim
V08 framework. Furthermore, since the CMB collocations already use actual
observations, ERA5 precipitation over sea ice is added directly to the CMB
collocations instead of creating. For non-GMI sensors, a separate
ERA5-collocation dataset is created that uses actual observations from the
target sensor.

\subsubsection{Finetuning}

For all non-GMI sensors except AMSRE, the training is conducted in two steps:
The basic training using the training data described above followed by a
finetuning step. The basic training produces the Day-1 retrieval, which can be
deployed as soon as calibrated observations from the target sensor become
available. For sensors for which at least several months of observations are
available and that overlap with GMI or another suitable reference sensor, an
additional finetuning is performed. The finetuning uses collocations of actual
observations from the target sensor and GPROF V08 retrievals from a reference
sensor. GMI is used as reference sensor for all sensors that overlap with it,
while collocations with SSMIS retrievals (themselves finetuned using GMI
reference estimates) were used for SSMI and AMSU-B. The GMI-based finetuning
also use the ERA5 collocations from the basic training because GMI finetuning
collocations are limited to observations equator-ward of 70 degrees and thus
fail to capture some surface processes occurring in the polar regions (such as
the formation of melt ponds). The only sensor for which no suitable reference
could be found was AMSR-E. While AMSR-E overlaps with TMI, it was decided not to
use it as a reference sensor due to its limited latitudinal coverage. Instead,
the AMSR-E retrieval uses the finetuned AMSR2 retrieval model, which was found
to yield slightly higher accuracy than the simulation-based Day-1 retrieval.

\subsubsection{Ancillary Data}

The GPROF algorithm traditionally uses ancillary data from numerical weather
forecasts and analyses as well as other sources to help constrain the
precipitation retrieval. In order to satisfy real-time processing requirements,
GPROF can be run with different ancillary-data configurations. For near-real
time (NRT) operations, the environmental data is sourced from operational
forecasts from the Japanese Meteorological Agency (JMA). The standard (STD)
configuration, which is run once full orbits are available, uses ancillary data
from JMA’s global analysis. Finally, the climate version (CLI) of GPROF uses
ancillary data from ERA5. GPROF V08 inherits many of the ancillary variables from
earlier versions of the algorithm. The principal change is that instead of using
discrete surface types that are calculated using a combination of static and
dynamic variables, the GPROF V08 retrieval ingests the underlying variables directly.

Because the availability of ancillary variables differs between the JMA
forecast and analysis and the ERA5 reanalysis, GPROF V08 uses multiple ancillary data
configurations. The ancillary variables and their use within each ancillary data
configuration are listed in table~\ref{tab:ancillary_data}. The land fraction
for each sensor is calculated from the MODIS land-sea mask and takes into
account the approximate footprint of the sensor on the Earth surface. The
mountain type classification is based on \citet{karagulle_2017}. The surface
elevation is derived from \citet{danielson_2011}. Two-meter temperature,
total-column water vapor, orographic wind, ten-meter wind are all derived from
the JMA NWP and analysis data for near-real time and standard operations and
from ERA5 for climate operations. Sea ice fraction and snow mask in near-real
time and standard operations are derived from AutoSnow \citep{romanov_2017}. In
climate operations, the ERA5 snow depth is used instead of the Autosnow-based snow
mask and the sea ice fraction is derived from ERA5.

To avoid having to train separate models for all sensors and ancillary-data
configurations, a configuration is chosen randomly for each training sample. In
order to support running the retrieval without ancillary data, all ancillary
variables are dropped with a similar probability as any other configuration. As
will be shown below, running the retrieval without ancillary data yields almost
identical accuracy as the retrieval with the most sophisticated ancillary
configuration. However, it substantially simplifies application of the retrieval
since it doesn’t require external datasets or the GPROF preprocessor software,
which combines the brightness temperatures and the ancillary data. With the
no-ancillary data (NA) retrieval mode, GPROF V08 is the first GPROF
implementation that can be directly run on GPM L1C files and thus is truly
openly accessible to the community.

\begin{table}[hbpt]
  \caption{
    List of variables included in the ancillary data used by GPROF V08. The first column lists the variable name. Column two, three, and four list which variables are included in the Near-Real Time, Standard, and Climatology ancillary-data
    configurations, respectively.
  }
  \label{tab:ancillary_data}
  \begin{tabular}{l|ccc}
    Variable & Near-Real Time (NRT) & Standard (STD) & Climatology (CLI) \\
    \hline
    Land fraction               & $\checkmark$ & $\checkmark$ & $\checkmark$ \\
    Mountain Type               & $\checkmark$ & $\checkmark$ & $\checkmark$ \\
    Elevation                   & $\checkmark$ & $\checkmark$ & $\checkmark$ \\
    Leaf Area Index Climatology & $\checkmark$ & $\checkmark$ & $\checkmark$ \\
    Two-Meter Temperature       &              & $\checkmark$ & $\checkmark$ \\
    Total Column Water Vapor    &              & $\checkmark$ & $\checkmark$ \\
    Orographic Wind             &              & $\checkmark$ & $\checkmark$ \\
    Ten-Meter Wind              &              & $\checkmark$ & $\checkmark$ \\
    Moisture Convergence        &              & $\checkmark$ & $\checkmark$ \\
    Convective Precipitation    &              &              & $\checkmark$ \\
    Sea-Ice Fraction            & $\checkmark$ & $\checkmark$ & $\checkmark$ \\
    Snow Depth                  &              &              & $\checkmark$ \\
    Snow Mask                   & $\checkmark$ & $\checkmark$ &              \\
  \end{tabular}
\end{table}

\subsubsection{Data Augmentation}

A consequence of the GPROF-NN 3D retrieval operating on image patches is that
observation artifacts such as radio frequency interference (RFI) or missing scan
lines can impact pixels outside the directly affected area. This is problematic
since input observation quality flags operate pixel-wise, making it difficult to
mask retrieval output in the vicinity of corrupted input pixels. While the fine
tuning hardens the retrieval against commonly occurring artifacts such as RFI,
it typically uses only observations from one sensor instance of each sensor type
which may not fully reflect the range of artifacts occurring across all
instances of a sensor type. Moreover, since the 2BCMB product does not produce
precipitation estimates for pixels affected by RFI, the original GMI training
data does not constrain precipitation estimates for pixels affected by RFI.

To harden the retrievals against commonly occurring observation artifacts, both
training stages apply the following random augmentations to each training scene:

\begin{enumerate}
\item Load a random ancillary data configuration or drop all ancillary variables.
\item Drop each channel with a probability of 10$\%$.
\item Drop or corrupt 1 – 30 randomly selected scanlines. The corruption sets
the scanlines to a random brightness temperature between 310 and 350 K emulating
observation artifacts observed for one of MHS sensors.
\item Observations from each channel are randomly set to missing with a chance
of 10 \%. For GMI, random Gaussian blobs with a channel-wise randomized
amplitude of up 200 K are added to the channels with frequencies below 85 GHz to
simulate RFI.
\end{enumerate}

\subsubsection{Training}

All retrieval models are trained using the AdamW
\citep{loshchilov_decoupled_2019} optimizer for 220 epochs with warm restarts
\citep{loshchilov2017sgdrstochasticgradientdescent} after 20, 60, and 140
epochs. For the last 80 epochs only the fully connected output layers of the
surface precipitation estimates are trained while the remaining parameters of
the model remain frozen. The last training step was found to slightly improve
the representation of heavy precipitation rates. The fine tuning phase of the
training is run for 160 epochs with a warm restart after 80 epochs and the
freezing of all layers except the surface precipitation head for the last 80
epochs.

\subsection{Validation Data}

The principal aim of this study is to validate the novel GPROF V08 retrieval
and assess the improvements compared to the previous operational version GPROF
V07. In order to provide a comprehensive assessment across the full GPM
constellation, the validation makes use of independent precipitation estimates
from various sources, which we described below.

\subsubsection{The SatRain Benchmark Dataset}

To establish the basic accuracy of the GPROF V08 retrieval for GMI and ATMS in a
reproducible manner, we use the testing data from the SatRain benchmark dataset
\citep{pfreundschuh2026benchmark}. The testing data is derived from
gauge-corrected ground-based radar measurements over CONUS and the Korean
peninsula, as well as gauge measurements from the WegenerNet network in Austria.
The dataset provides two subsets consisting of overpasses from either GMI or
ATMS. We follow the standard SatRain evaluation protocol, which evaluates all
retrievals on a regular latitude-longitude grid with a resolution of
$0.036^\circ$. Over CONUS the evaluation is limited to pixels for which the
radar-quality index exceeds 0.5.

\subsubsection{Additional Ground-Based Radar Data}

Since the SatRain dataset is limited to GMI and ATMS overpasses, we create
additional independent validation datasets for AMSR2, SSMIS on F17, ATMS on
NOAA20, and MHS on NOAA19 using gauge-corrected ground-based radar estimates
from MRMS. The validation datasets for these sensors are created using all
available overpasses over the MRMS domain from 2022. The evaluation follows the
SatRain protocol evaluating all retrievals on a regular latitude-longitude grid
with a resolution of $0.036^\circ$ and a minimum radar-quality index of $0.5$.

\subsubsection{GPM 2BCMB Collocations}

To provide additional validation of the GPROF V08 retrieval over ocean, we
extract collocations of the target sensors with the 2BCMB product using all
coincident overpasses during the year 2022. Since the 2BCMB retrievals are used
in the basic training of the GPROF V08 sensors, this data is not completely
independent. However, the time period is independent of any data used to
training any of the GPM sensors. Moreover, since 2BCMB precipitation fields are
the most accurate precipitation estimates available over global oceans, good
agreement with 2BCMB is arguably the best possible accuracy indicator available
over ocean. The evaluation uses the same protocol as the SatRain dataset
evaluating precipitation estimates on a regular latitude-longitude grid with a
resolution of $0.036^\circ$.

\subsubsection{TRMM 2BCMB Collocations}

Finally, in order to also validate retrievals from the pre-GPM era, validation
datasets for TMI, SSMI on F15, AMSU-B on NOAA-15, and AMSR-E are created using
reference precipitation estimates from the TRMM combined radar and radiometer
retrievals \citep{trmm_cmb_data} using all available collocations from the year
2004. The same dataset is used to validate retrievals over ocean and over land.

\subsection{Metrics}

We will use the following metrics to quantitatively assess the accuracy
of the GPROF V08 retrievals. Let $\hat{y}_i$ denote the retrieved precipitation
and $y_i$ the corresponding reference value for each validation sample $i = 1, \dots, N$.
We will use the relative bias in percent to measure systematic errors in the retrieval, with a
positive bias corresponding overestimation of precipitation and a negative bias
corresponding to underestimation.
\begin{align}
  \mathrm{Bias} &= 100 \cdot \sum_{i=1}^{N}\frac{(\hat{y}_i-y_i)}{y_i}
\end{align}

To assess instantaneous retrieval errors we will us the mean absolute error (MAE),
mean-squared error (MSE), and symmetric mean percentage error (SMAPE):
\begin{align}
  \mathrm{MAE} &= \frac{1}{N}\sum_{i=1}^{N}\left|\hat{y}_i-y_i\right| \\
  \mathrm{MSE} &= \frac{1}{N}\sum_{i=1}^{N}(\hat{y}_i-y_i)^2 \\
  \mathrm{SMAPE} &= \frac{1}{N}\sum_{i, 0.1 < y}^{N}\frac{|\hat{y}_i-y_i|}{0.5 \cdot (|\hat{y}_i| + |y_i|)}
\end{align}

Additionally, we will use the linear correlation coefficient to assess the covariability between
retrieved and reference precipitation estimates.

\begin{align}
  \text{Correlation Coef.} &= \frac{\sum_{i=1}^{N}(\hat{y}_i-\bar{\hat{y}})(y_i-\bar{y})}{\sqrt{\sum_{i=1}^{N}(\hat{y}_i-\bar{\hat{y}})^2 \sum_{i=1}^{N}(y_i-\bar{y})^2}}.
  \label{eq:metrics_end}
\end{align}

\subsubsection{Effective Resolution}

To complement the point-wise error metrics defined above, we will use the
effective resolution to assess the capability of the retrieval to capture
precipitation features across spatial scales. The effective resolution
represents the smallest scale of precipitation features that a retrieval is
able to capture on average. We follow the formulation introduced in
\citet{pfreundschuh_gprof_2024}, which is based on the work by
\citet{guilloteau_multiscale_2020} and \citet{guilloteau2021well}.
The effective resolution is defined in terms of the spectral coherence
$C_{\hat{y},y}$ between the retrieved precipitation field
$\hat{y}$ and and reference precipitation field $y$:

\begin{align}
C_{\hat{y},y}(k) &= \frac{\left|P_{\hat{y},y}(k)\right|} {\sqrt{P_{\hat{y}}(k)P_y(k)}},
\end{align}

where \(P_{\hat{y}}\), \(P_y\), and \(P_{\hat{y},y}\) are the radially averaged auto- and
cross-power spectral densities
\begin{align}
P_{\hat{y}}(k) &= \left\langle  \text{DCT}(\hat{y})(k_x,k_y) \cdot \text{DCT}(\hat{y})(k_x,k_y)^{*}\right\rangle, \\
P_{y}(k) &= \left\langle \text{DCT}(y)(k_x,k_y) \cdot \text{DCT}(y)(k_x,k_y)^{*}\right\rangle, \\
P_{\hat{y},y}(k) &= \left\langle  \text{DCT}(\hat{y})(k_x,k_y) \cdot \text{DCT}(y)h(k_x,k_y)^{*}\right\rangle.
\end{align}

Here $\text{DCT}(\cdot)$ are the coefficients of a 2D discrete cosine transform
of the retrieved or reference precipitation field. The averages
$\langle\cdot\rangle$ are computed over non-overlapping windows of size
$32\ \text{px} \times 32\ \text{px}$ containing no invalid retrieved or
reference precipitation estimates and over all 2D DCT coefficients with the same
radial wavenumber. The effective resolution corresponds to half the wavelength of the finest
wavenumber $k$ at which the spectral coherence exceeds $\frac{1}{\sqrt{2}}$.

\section{Results}

\subsection{Test-Set Accuracy}

To establish the nominal accuracy of the GPROF V08 retrieval for GMI, we
evaluate the GPROF V08 (GMI) retrieval model on the independent test data which
consists of the training collocations from the first three days of every month
from October 2018 until October 2019. We restrict this analysis to GMI because
its training and testing data are based on actual observations and thus similar
to the conditions encountered in operational use. For non-GMI sensors, the basic
training uses simulated observations, and performance on the corresponding
testing data therefore does not provide a representative estimate of operational
retrieval accuracy.

\begin{figure}[hbpt]
  \centering
  \includegraphics[width=\textwidth]{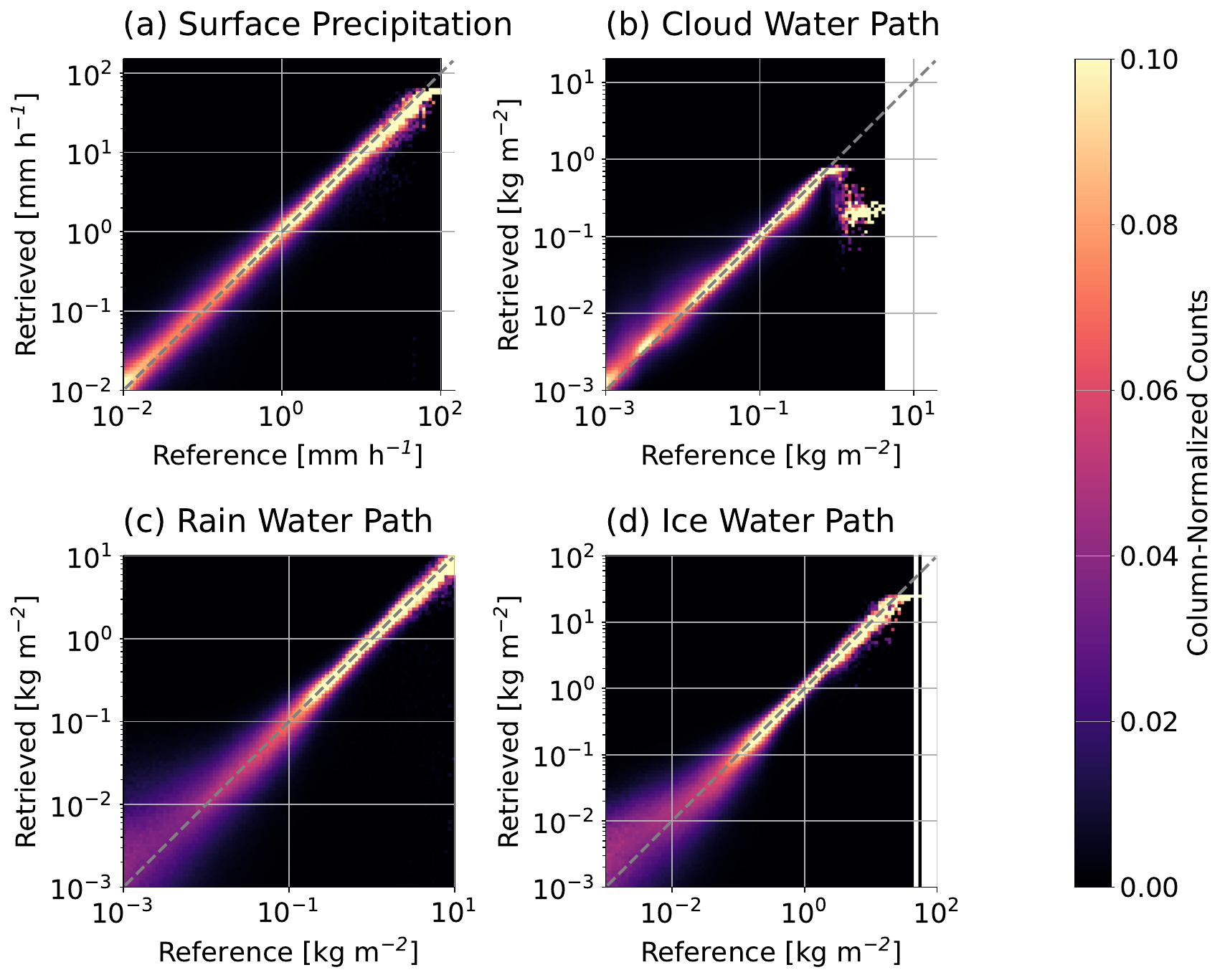}
  \caption{
    Scatter plots of reference and GPROF V08 (GMI) retrieved
surface precipitation (a), cloud water path (b), rain water path (c), and ice
water path (d) calculated using the testing data. Counts have been normalized
column-wise to improve visibility.
  }
  \label{fig:test_results_scatter}
\end{figure}

Figure~\ref{fig:test_results_scatter} shows scatter plots comparing the
retrieved posterior mean of GPROF single-level output variables with the
corresponding reference values. The retrieved values generally follow the
one-to-one line but exhibit slight tendencies to overestimation at the low end
and underestimation at the high end of the target distributions. Surface
precipitation, cloud water path, ice water path, also exhibit a visible cap in
the retrieved values. This behavior is likely due to the saturation of the
signal in the input observations at high intensities of the target variable,
which limits the retrieval's to resolve the corresponding reference-data
variability.

\begin{figure}[hbpt]
  \centering
  \includegraphics[width=0.5\textwidth]{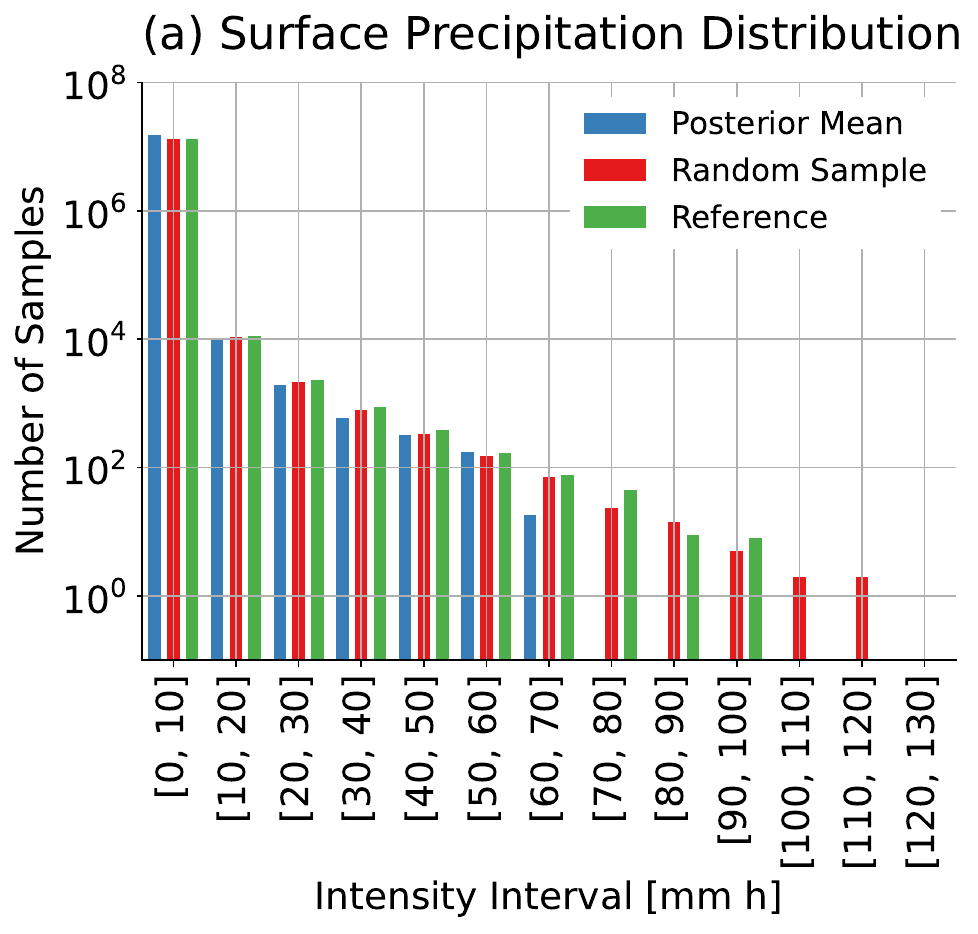}
  \caption{
    Distributions of retrieved and reference surface precipitation for the GPROF
V08 (GMI) retrieval. The graph compares the number of samples falling into 13
regularly-spaced precipitation intensity bins between $0$ and
$130\ \unit{mm\ h^{-1}}$ when the retrieved posterior distribution is used to
calculate the posterior mean (blue bars) or draw a random sample (red bars) to
the distribution of reference precipitation rates (green bars).
  }
  \label{fig:test_distributions}
\end{figure}

Evidence for this claim comes from the distribution retrieved precipitation
intensities shown in Fig.~\ref{fig:test_distributions}. The figure compares the
distributions of the posterior mean and random samples drawn from the retrieval
posterior distribution. The random samples recover and even slightly
overestimate the heavy tail of the reference precipitation distribution. This
indicates that the model has learned to identify situations in which very high
intensities occurs but that the observations do not provide sufficient
information resolve the variability in the reference data forcing it to resort
to reporting the conditional mean, which is less than the highest reference
precipitation rates. The overestimation occurring for the random samples is
likely due to the extrapolation of the cumulative distribution outside of the
lowest and highest predicted quantiles that is required to draw samples from the
CDF. The extrapolation assumes that the probability distribution function
linearly decreases to zero, which seems to exaggerate the tail of the
distribution. However, it should also be kept in mind that the reference data
itself is subject to non-negligible uncertainties particularly at these very
high intensities and that random error in the reference data contributed to the
non-resolved variability at high precipitation rates.

Table~\ref{tab:evaluation_metrics} summarizes the test-set accuracy metrics for
the principal output variables produced by the GPROF V08 retrieval. Relative
biases are within a few percent for most variables, while the retrievals
generally exhibit low errors and strong correlations with the reference values.
The accuracy metrics are generally slightly higher than those reported for the
original GPROF-NN prototype \citep{pfreundschuh_gprof-nn_2022}. Since the
evaluation presented here use similar testing datasets (up to the inclusion of
light precipitation from the GPROF-NN XPR retrieval) the results are generally
comparable. The operational GPROF V08 retrieval thus modestly improves upon the
nominal accuracy of the originally proposed GPROF-NN 3D retrieval. We attribute
this to the improved neural network architecture as well as the prediction of
non-transformed target variables instead of the log-linear transform used by the
original GPROF-NN 3D retrieval.

\begin{table}[ht]
  \centering
  \caption{Evaluation metrics for all retrieved quantities for the GPROF V08 (GMI) retrieval
    on the test data set.
  }
  \label{tab:evaluation_metrics}
  \begin{tabular}{lrrrrr}
    \hline
    Quantity & Bias [\%] & MAE [$\unit{mm\ h^{-1}}$] & MSE [$(\unit{mm\ h^{-1}})^2$] & SMAPE [$\unit{\%}$] & Corr. Coef. \\
    \hline
    Surface Precipitation & -1.17 & 0.039 & 0.07384 & 46.27  & 0.94 \\
    Cloud Water Path & 0.79 & 0.013 & 0.00096 & 36.10  & 0.94 \\
    Rain Water Path & -1.37 & 0.007 & 0.00235 & 62.68  & 0.96 \\
    Ice Water Path & -0.14 & 0.007 & 0.00247 & 63.80  & 0.96 \\
    Cloud Water Content & 0.61 & 0.001 & 0.00003 & 40.15  & 0.92 \\
    Rain Water Content & -1.65 & 0.001 & 0.00008 & 63.04  & 0.92 \\
    Snow Water Content & -0.77 & 0.001 & 0.00007 & 69.73  & 0.93 \\
    Latent Heating & -2.51 & 0.013 & 0.03590 & 74.38  & 0.81 \\
    \hline
  \end{tabular}
\end{table}

\subsection{Independent Validation}

Although good test-set performance is necessary for an accurate precipitation
retrieval, it is not sufficient to establish the operational accuracy of the
GPROF V08 retrieval as it does not account for uncertainties in the reference
data. Moreover, since GPROF has to rely on synthetic observations for the
basic training of non-GMI sensors, the test-set error is not representative of the
error that has to be expected when applying the retrieval to actual
observations. In the following, we therefore asses the GPROF V08 retrieval for
GMI and all principal GPM constellation sensors that were operational prior to
2024 using precipitation independent measurement sources as well as combined
radar and radiometer retrievals from GPM and TRMM, where necessary.

GPROF V08 can be run using several ancillary data configurations depending on
the operational context. In the evaluation below, we assess the accuracy of
the no-ancillary (NA), near-real-time (NRT), and climate (CLI) ancillary data
configurations. The three configurations span the spectrum of ancillary-data
complexity with CLI representing the most comprehensive ancillary data and NA no
ancillary data at all. Since the standard configuration is very similar to the
CLI configuration, we do not assess it independently.

\subsubsection{SatRain}

\begin{figure}[hbpt]
  \centering
  \includegraphics[width=\textwidth]{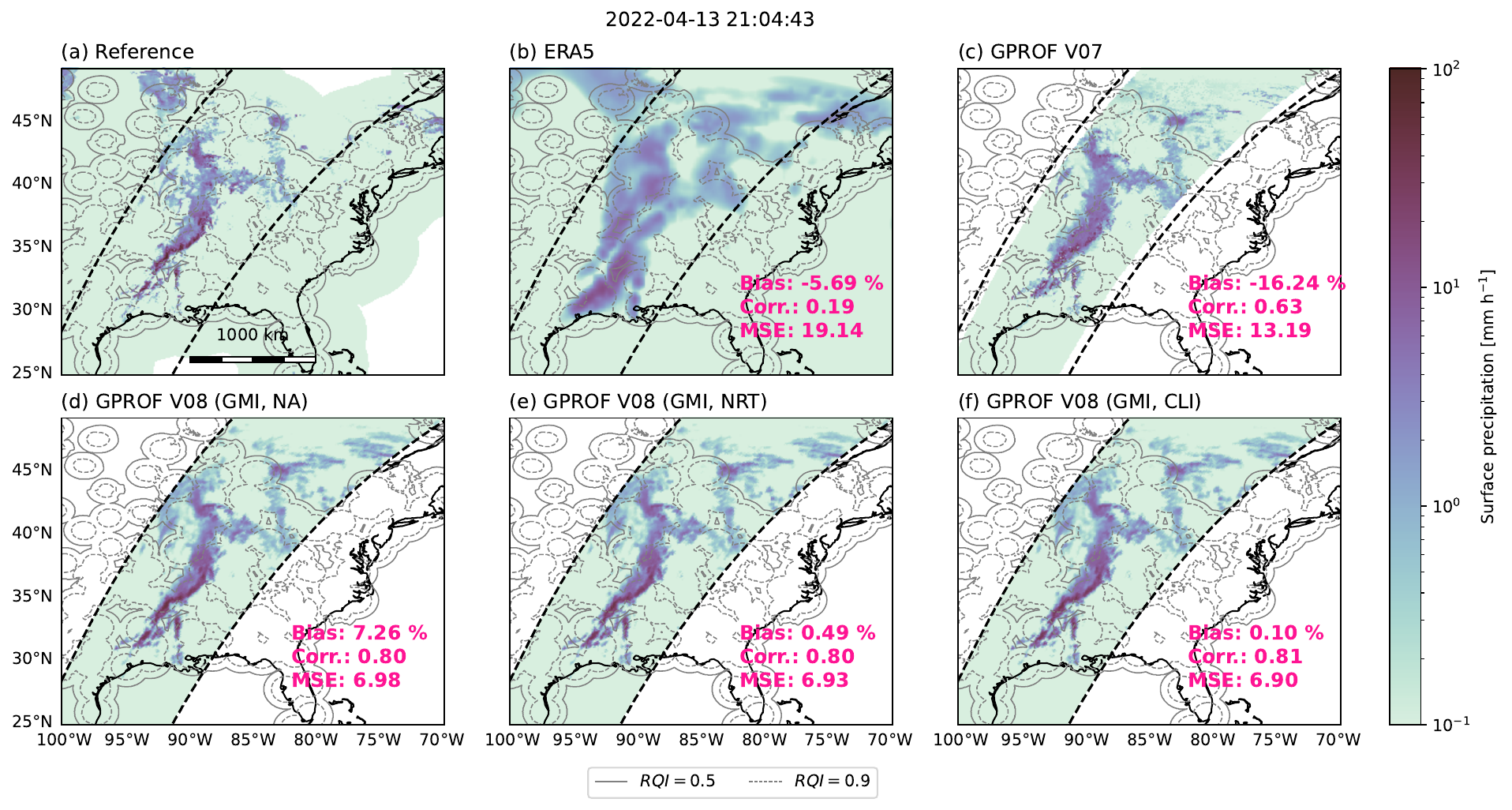}
  \caption{
Reference and retrieved precipitation for a convective storm system on April 13,
2022 over CONUS. Panel (a) shows reference precipitation estimates from
gauge-corrected ground-based radars from the SatRain benchmark dataset. Panels
(b) – (f) show the results from (b) ERA5, (c) GPROF V07 (GMI), and GPROF V08 (GMI)
 using no ancillary data (NA, (d)), using only static ancillary
data (NRT, (e)), and using all ancillary data (CLI, (f)).
  }
  \label{fig:satrain_case_study_gmi_1}
\end{figure}

To assess the performance of the GPROF V08 GMI retrieval for individual
precipitation events, we examine two case studies from the SatRain dataset
presented in \citet{pfreundschuh2026benchmark}.
Figure~\ref{fig:satrain_case_study_gmi_1} compares GPROF V08 retrieval results
with those from GPROF V07 and the ERA5 reanalysis for a convective storm system
observed over eastern CONUS on April 13, 2022. Although GPROF V07 already
performs well in this case, the GPROF V08 retrieval yields better agreement with
the MRMS reference estimates. Relative to GPROF V07, GPROF V08 reduces the magnitude
of the bias and the MSE while increasing the linear correlation coefficient. The
differences between the three ancillary data configurations remain small. The
NRT and CLI configurations exhibit the lowest bias magnitudes, whereas the MSE
and correlation coefficient differ by less than than 2\%.

\begin{figure}[hbpt]
  \centering
  \includegraphics[width=\textwidth]{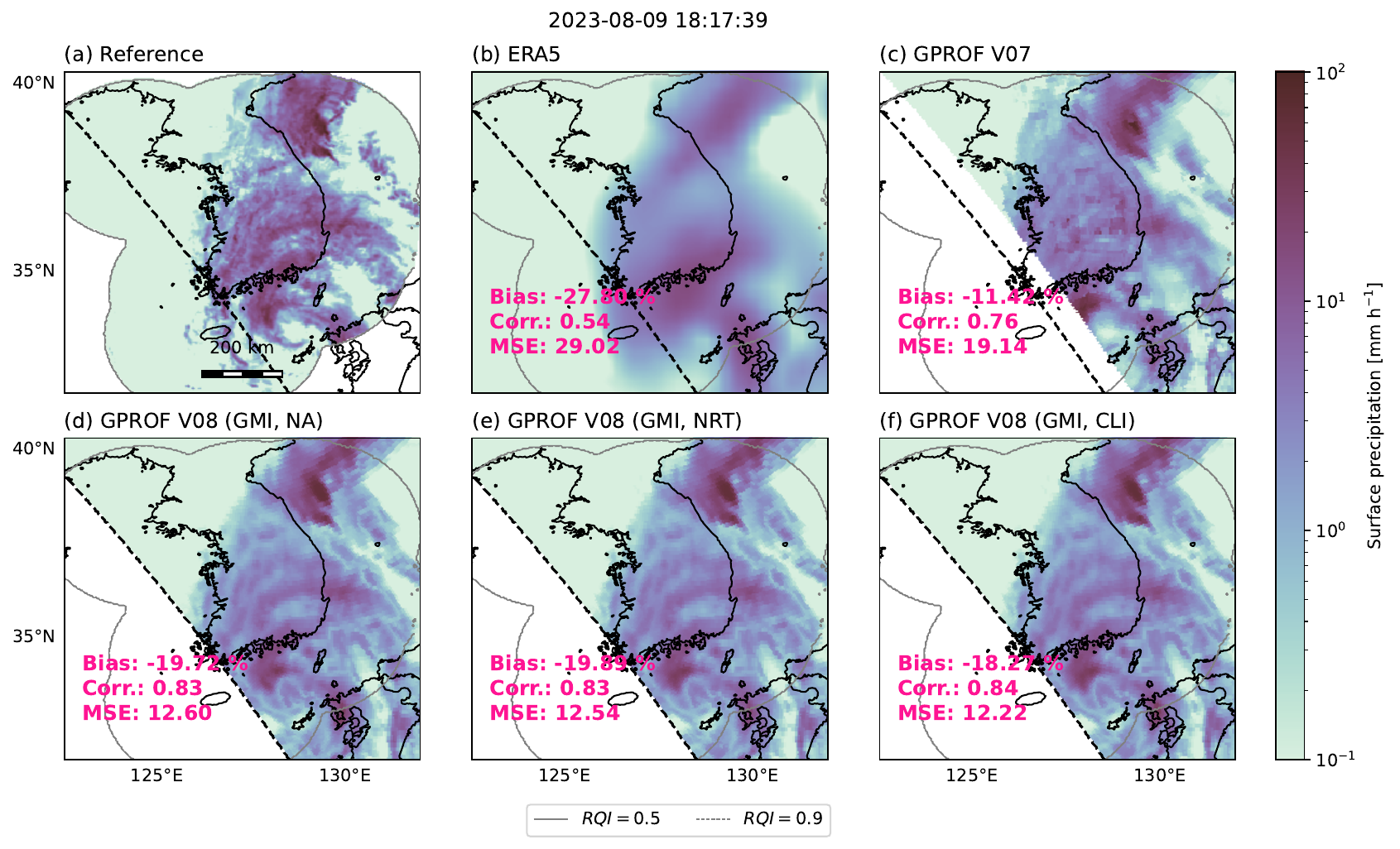}
  \caption{
    As Fig.~\ref{fig:satrain_case_study_gmi_1} but for the landfall of Typhoon Khanun on the Korean peninsula.
  }
  \label{fig:satrain_case_study_gmi_2}
\end{figure}

Figure~\ref{fig:satrain_case_study_gmi_2} presents the corresponding results for
a GMI overpass of Typhoon Khanun during its landfall over the Korean Peninsula.
The GPROF V08 retrievals exhibits a stronger negative precipitation bias than
GPROF V07 but achieve lower MSEs and higher correlations with the reference
estimates. This combination indicates that V08 reproduces the spatial
distribution of precipitation more accurately despite underestimating its
overall magnitude. The improvement is particularly apparent over land, where
GPROF V08 captures the typhoon’s rainband structure, which is largely absent
from the GPROF V07 retrieval. However,  the GPROF V08 retrievals
exhibit a tendency to overestimate light precipitation over oceans, which may be
a sideeffect of the light precipitation added by the GPROF-NN XPR retrieval.

\begin{figure}[hbpt]
  \centering
  \includegraphics[width=\textwidth]{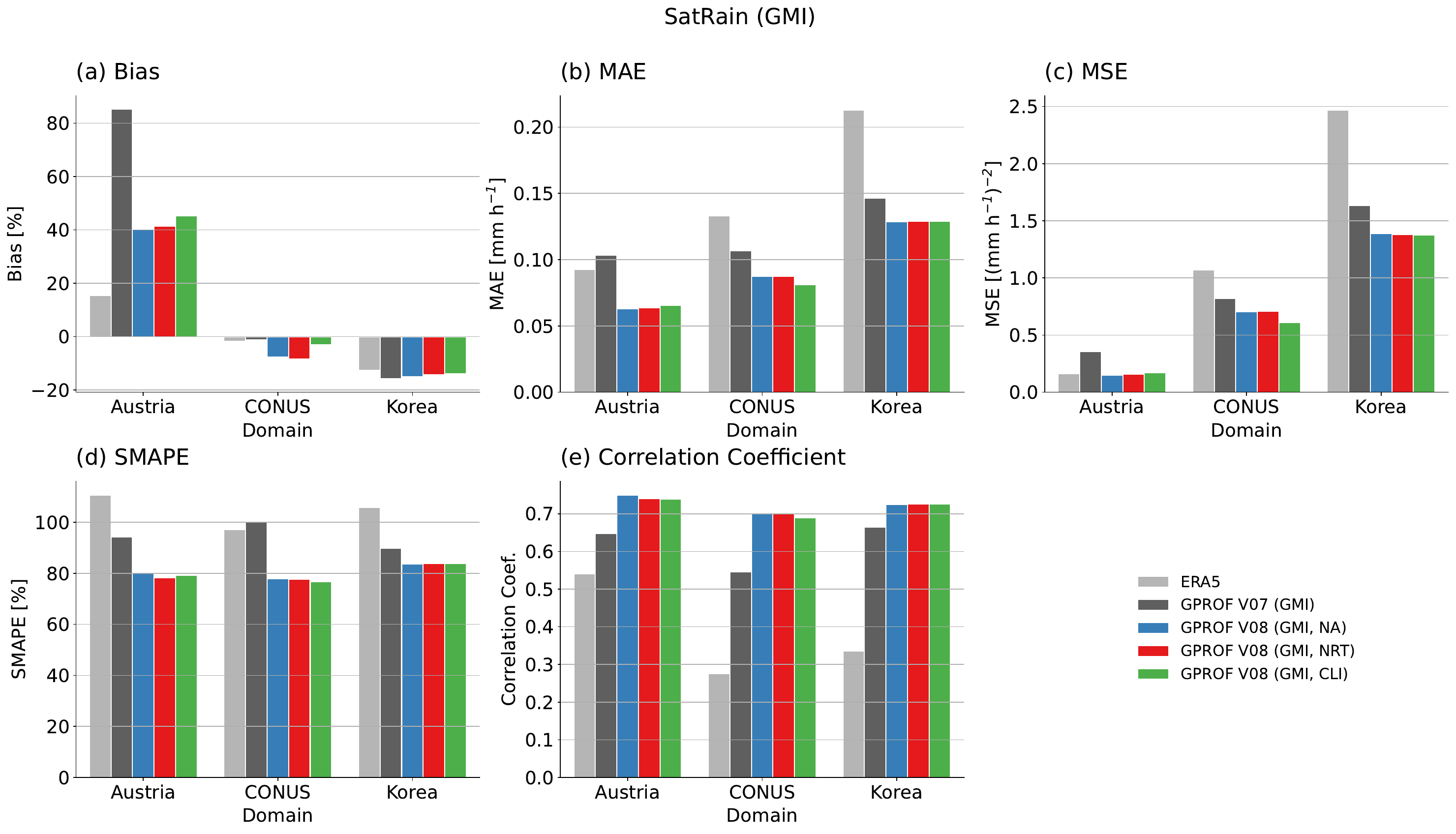}
  \caption{
    GPROF V08 (GMI) retrieval accuracy across the three domains of the SatRain
    testing dataset compared to ERA5 and GPROF V07 (GMI). Panel (a) shows the
    relative bias, panel (b) the mean absolute error (MAE), panel (c) the mean
    squared error (MSE), panel (d) the symmetric mean percentage error for reference
    precipitation estimates exceeding $0.1\ \unit{mm\ h^{-1}}$, and panel (e) the linear
    correlation coefficient.
  }
  \label{fig:satrain_stats_gmi}
\end{figure}

The error statistics for the full SatRain testing data for the GPROF V08 GMI
retrieval are displayed in Fig.~\ref{fig:satrain_stats_gmi}. In terms of overall
biases, all assessed precipitation datasets show similar tendencies. While all
estimates have very small biases over CONUS, they exhibit more systematic
underestimation over the Korea domain with biases ranging from -12\% to -15\%.
Over the Austria domain, all estimates exhibit notable positive biases ranging
from 15\% for ERA5 to 85\% for GPROF V07. The years 2021 and 2022 used for the
evaluation correspond to two particularly dry years in the WegenerNet record
suggesting that the remotely-sensed estimates may not be able to capture the
full extent of the regional climate variability. The biases for GPROF may be
further increased by the orographic enhancement factors that are applied to the
training data to account for the underestimation of orographic precipitation in
the 2BCMB reference data.

In terms of retrieval errors, the GPROF V08 retrieval for GMI reduces the MAE,
MSE, SMAPE, and increase the linear correlation coefficient compared to GPROF
V07 for GMI across all domains. The evaluation of the GPROF V08 GMI retrieval
thus confirms the findings from the two case studies above: The GPROF V08
retrieval for GMI achieves robust improvements compared to GPROF V07.
Furthermore, the ancillary data is found to have minimal impact on the accuracy
with the NA configuration achieving almost the same accuracy as the results of
the CLI configuration, which uses the most comprehensive selection of ancillary
data.

\begin{figure}[hbpt]
  \centering
  \includegraphics[width=\textwidth]{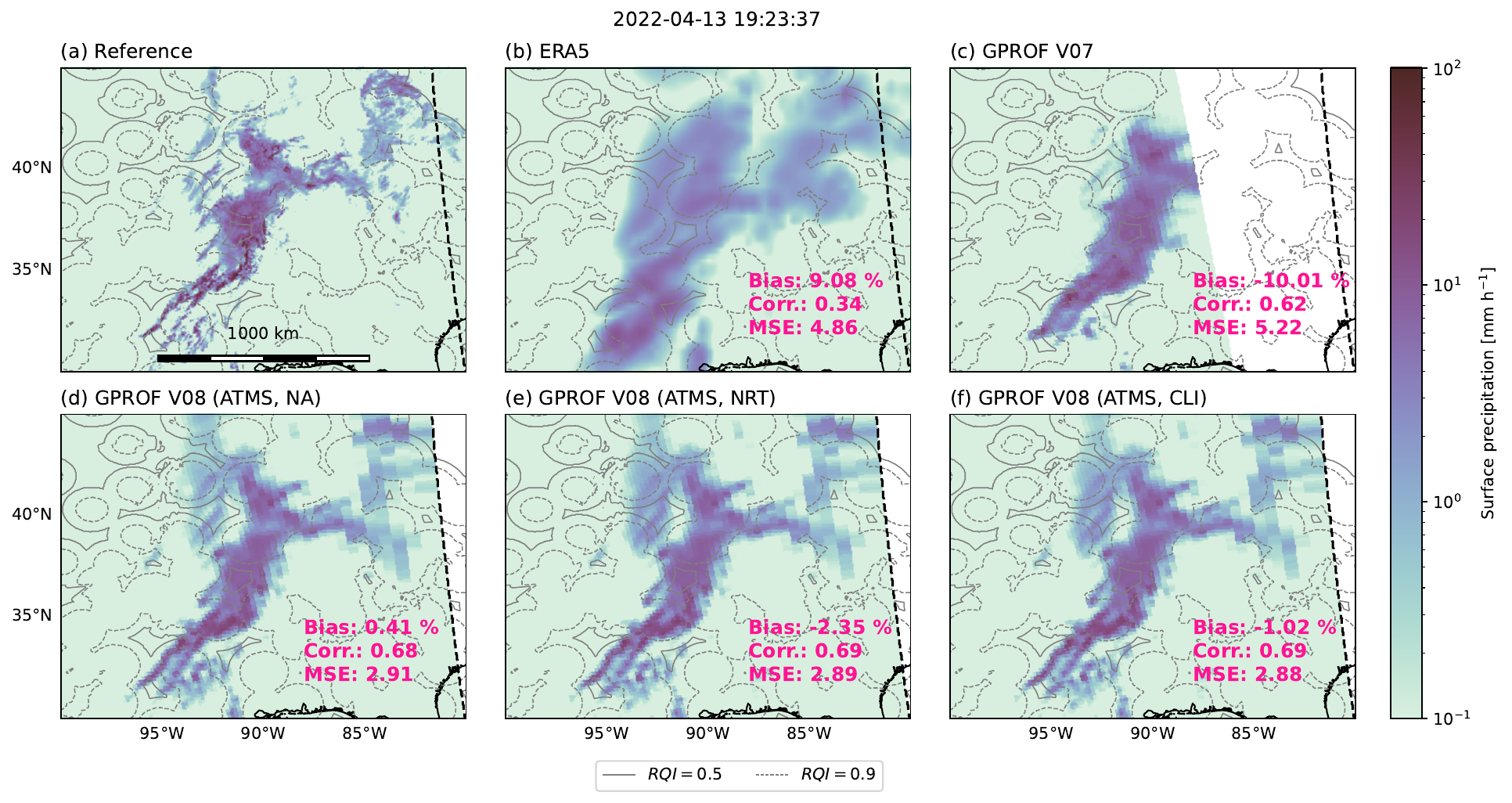}
  \caption{
  Reference and retrieved precipitation for an ATMS overpass over the convective
storm system shown Fig.~\ref{fig:satrain_case_study_gmi_1}. Panel (a) shows
reference precipitation estimates from gauge-corrected ground-based radars from
the SatRain benchmark dataset. Panels (b) – (e) show the results from ERA5,
GPROF V07 (ATMS) and GPROF V08 (ATMS) for the retrieval using no ancillary data (NA,
(c)), using only static ancillary data (NRT, (d)), and using all ancillary data
(CLI, (e)).
  }
  \label{fig:satrain_case_study_atms_1}
\end{figure}

To evaluate the GPROF V08 retrieval for ATMS, we consider overpasses close to
the two GMI case studies shown in Fig.~\ref{fig:satrain_case_study_gmi_1} and
Fig.~\ref{fig:satrain_case_study_gmi_2}. The results for the convective storm
over eastern CONUS are shown in Fig.\ref{fig:satrain_case_study_atms_1}. GPROF
V08 for ATMS shows visible improvements in its ability to resolve the convective
structures in the storm compared to GPROF V07. An additional difference between
GPROF V07 and GPROF V08 is that the retrieval is performed even for the outer
swath edges of the ATMS swath. Although the spatial resolution is visibly
degraded, the retrieval still produces reasonable results. The differences
between the three ancillary configurations remains small also for this case
study.

\begin{figure}[hbpt]
  \centering
  \includegraphics[width=\textwidth]{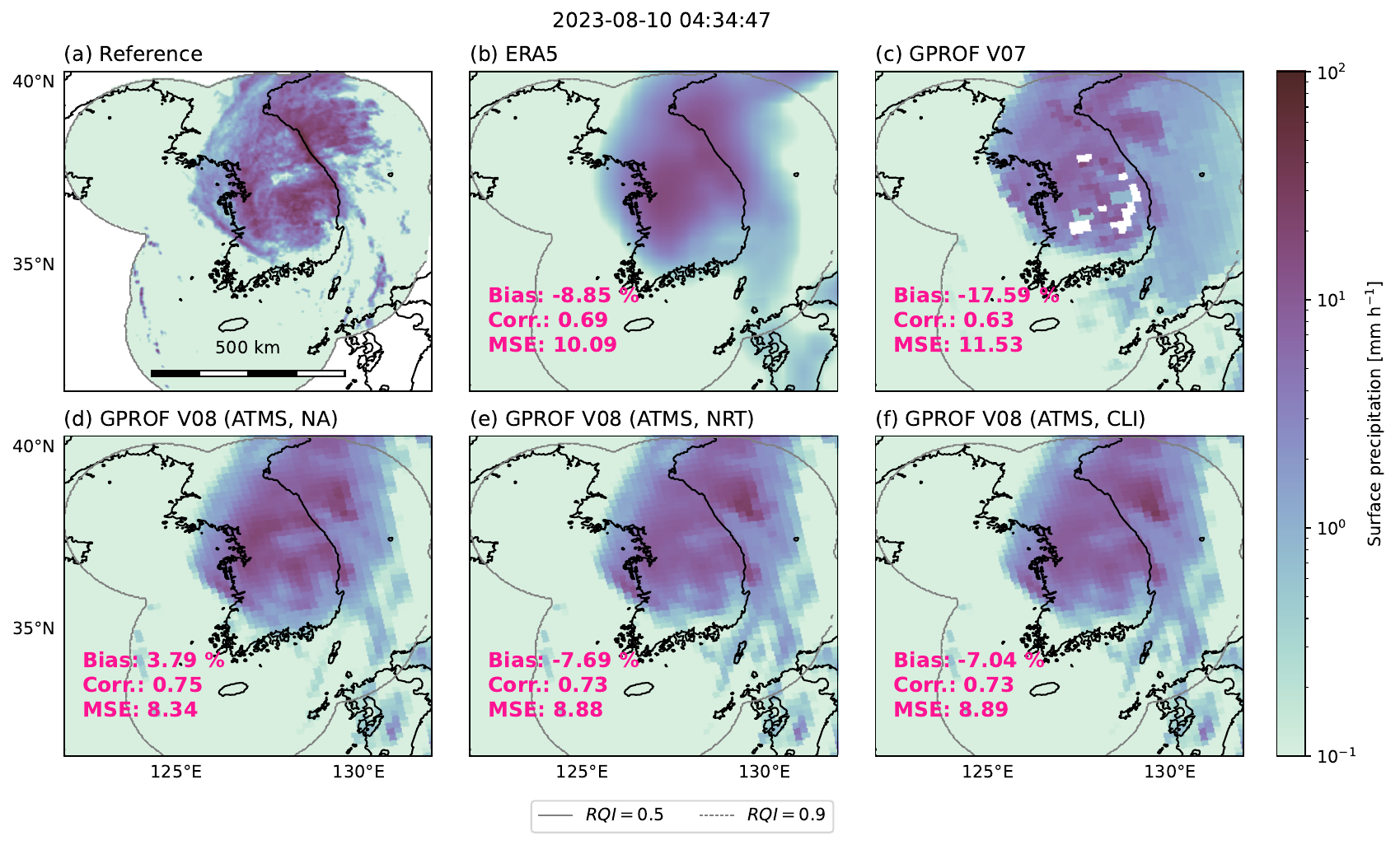}
  \caption{
  As Fig.~\ref{fig:satrain_case_study_atms_1} but for the landfall of Typhoon
Khanun on the Korean peninsula.
  }
  \label{fig:satrain_case_study_atms_2}
\end{figure}

Retrieval results for an ATMS overpass over Typhoon Khanun are shown in
Fig.~\ref{fig:satrain_case_study_atms_2}. Also for this case study, the GPROF
V08 retrievals reduce the MSE and increase the linear correlation. The GPROF V07
retrieval strongly overestimates the spatial extent of precipitation trailing
the Typhoon. Moreover, the retrieval reports several missing values over land
that are caused by the retrieval database not containing enough samples with
matching observations. This is because GPROF V07 introduced special mountain
classes, which in these particular cases cause the retrieval to fail because of
each class contains too few samples to match the input observations.

\begin{figure}[hbpt]
  \centering
  \includegraphics[width=\textwidth]{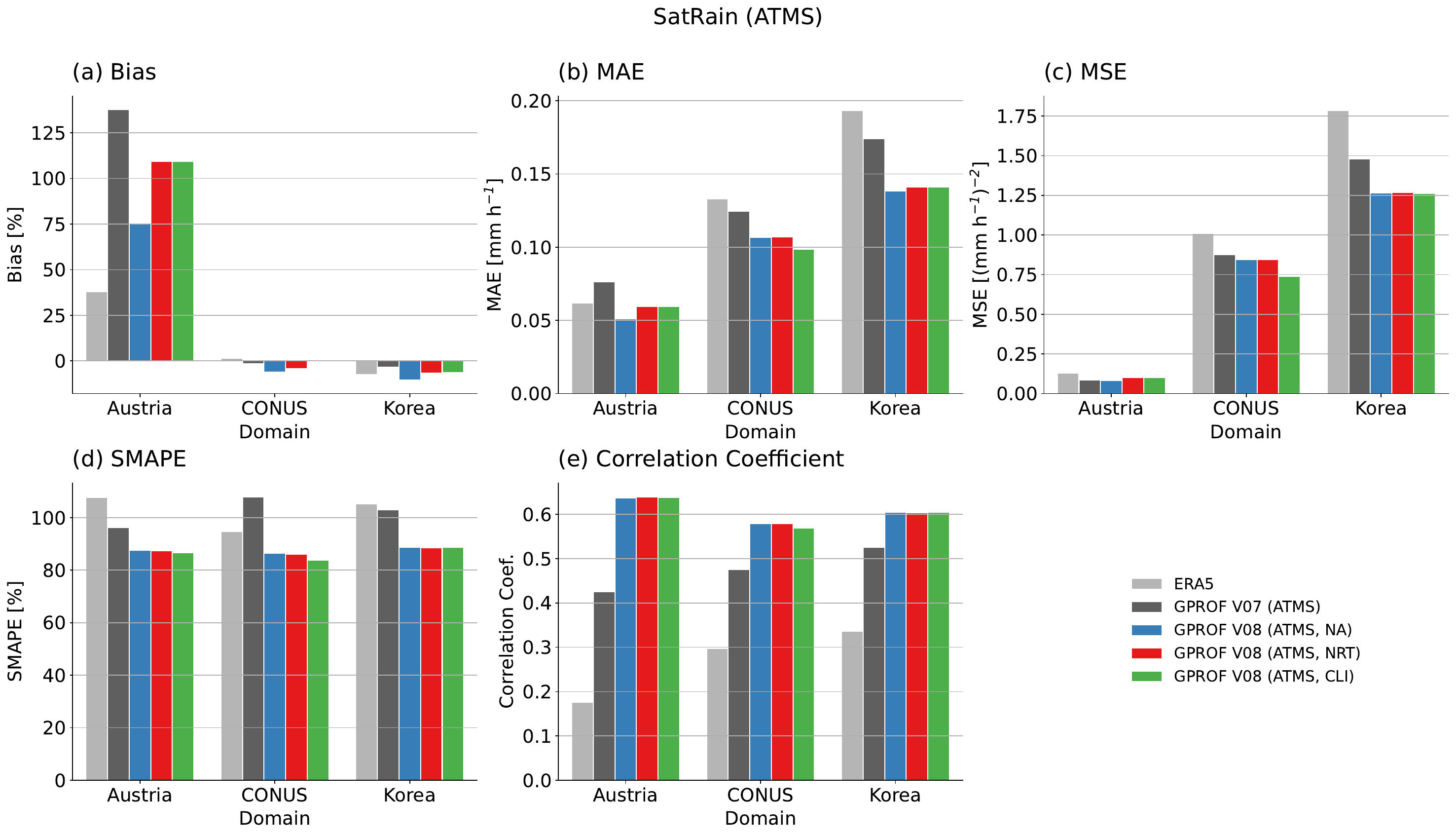}
  \caption{
    GPROF V08 (ATMS) retrieval accuracy across the three domains of the SatRain
    testing dataset compared to ERA5 and GPROF V07 (ATMS). Panel (a) shows the
    relative bias, panel (b) the mean absolute error (MAE), panel (c) the mean
    squared error (MSE), panel (d) the symmetric mean percentage error for reference
    precipitation estimates exceeding $0.1\ \unit{mm h^{-1}}$, and panel (e) the linear
    correlation coefficient.
  }
  \label{fig:satrain_stats_atms}
\end{figure}

Accuracy metrics for the whole SatRain testing data for ATMS are displayed in
Fig.~\ref{fig:satrain_stats_atms}. The biases generally follow the same pattern
as GMI with strong overestimation over the Austria domain, slightly negative
biases over the Korea domain, and close to no bias over CONUS. GPROF V08 for
ATMS reduces the MAE, MSE, and SMAPE and increases the correlation across all
domains compared to V07. Notably, the differences between ancillary-data
configurations remain negligible even for ATMS despite its observations
being limited to only five channels with coarser resolution than those of GMI.

\clearpage
\subsubsection{GPM-Era Sensors}

To extend the validation against independent precipitation estimates to other
GPM sensors as well as precipitation measurements over ocean, we validate
retrievals from AMSR2, ATMS, MHS, and SSMIS against gauge-corrected
precipitation estimates from ground-based radar over CONUS and combined radar
and radiometer retrievals from the GPM 2BCMB product. The operational retrieval
models for all non-GMI sensors have gone through the finetuning stage that uses
actual observations collocated with retrieval results from a reference sensors.
To assess the effect of the finetuning, this analysis also includes the results
of the Day-1 retrieval, which is the result from the basic training that is
largely based on synthetic observations.

\begin{figure}[hbpt]
  \centering
  \includegraphics[width=\textwidth]{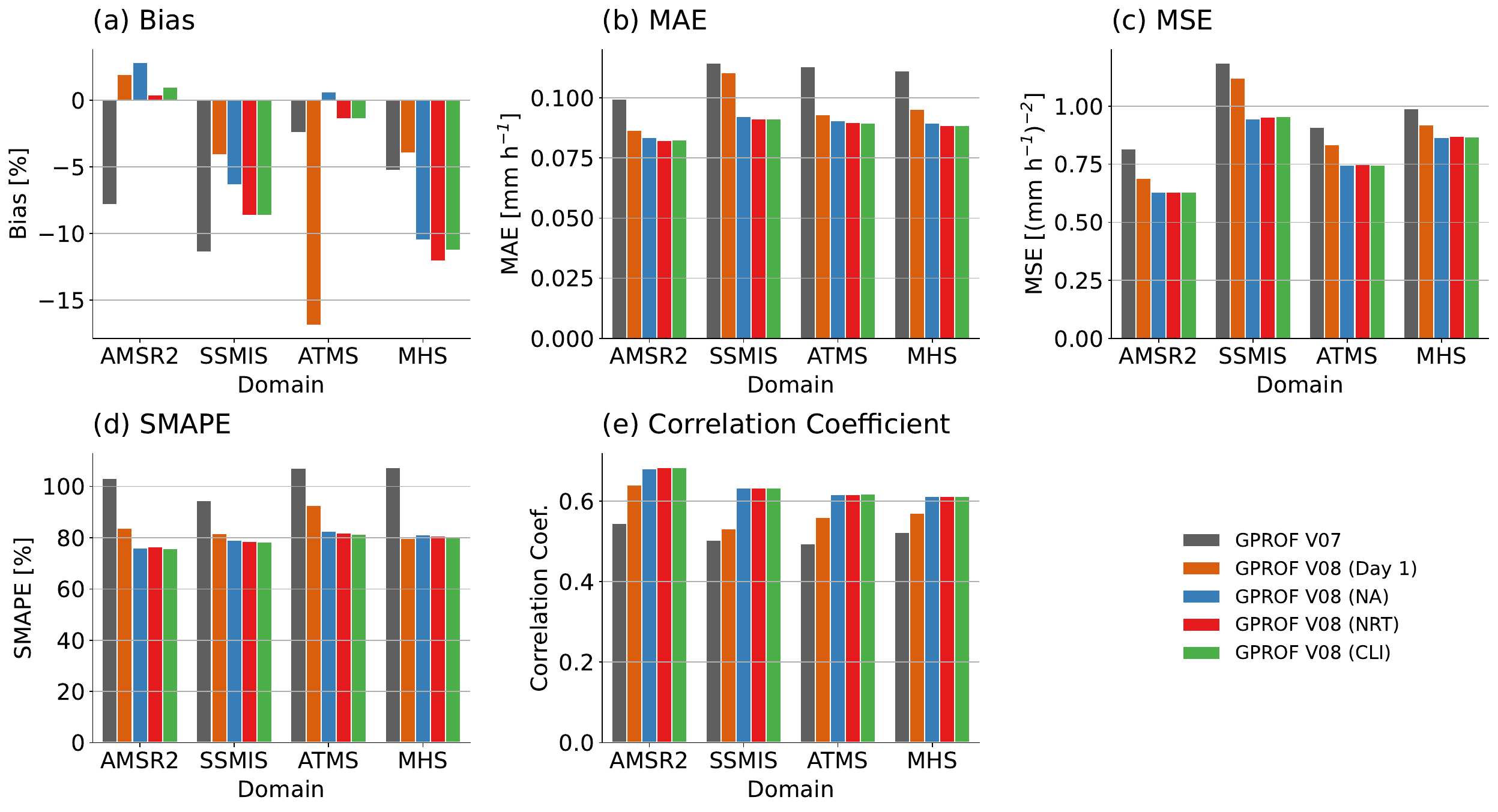}
    \caption{
      Retrieval accuracy of GPROF V08 for principal GPM-era constellation
sensors with respect to MRMS gauge-corrected ground-based radar measurements.
  }
  \label{fig:validation_conus_gpm}
\end{figure}

The accuracy metrics for all MRMS overpasses during 2022 are shown in
Fig.\ref{fig:validation_conus_gpm}. The retrieval biases vary strongly be
sensors and retrieval algorithm. The GPROF V08 Day-1 retrieval for ATMS exhibits
the strongest low bias for ATMS but remains within $5\%$ of MRMS for thee other
sensors. GPROF V07 shows a general tendency to underestimation with biases
ranging from $-2.4\%$ for ATMS and $-11.4\%$ for SSMIS. The biases of the
finetuned GPROF V08 retrievals range from $-12\%$ for MHS to $+2.8 \%$ for
AMSR2. The GPROF V08 retrieval biases are notably higher for SSMIS and MHS than
they are for AMSR2 and ATMS. It was found during development that the
neural-network-based GPROF V08 retrieval is sensitive to artifacts and
degradation in the PMW observations that can cause similar climatological biases
without affecting the instantaneous retrieval accuracy notably. However, these
systematic errors are also sensitive to climatological differences between the
training targets and the validation data \citep{pfreundschuh_gprof_2024}.
Uniquely attributed a cause to these biases will therefore require further
investigation.

In terms of the retrieval errors, the finetuned GPROF V08 retrievals show
notable improvements across all sensors compared to both GPROF V07 as well as
the Day-1 retrievals. The Improvements of the finetuned GPROF V08 retrievals
compared to GPROF V07 range from 16\% to 26\% across MAE, MSE, SMAPE, and linear
correlation coefficients and all sensors considered here. While the Day-1
retrieval yields lower instantaneous errors and higher correlation than GPROF
V07, the finetuning adds further improvements for all sensors. The differences
between the ancillary data configurations remain small for all assessed sensors,
providing further evidence that ancillary data only contributes weakly to the
retrieval accuracy.

\begin{figure}[hbpt]
  \centering
  \includegraphics[width=\textwidth]{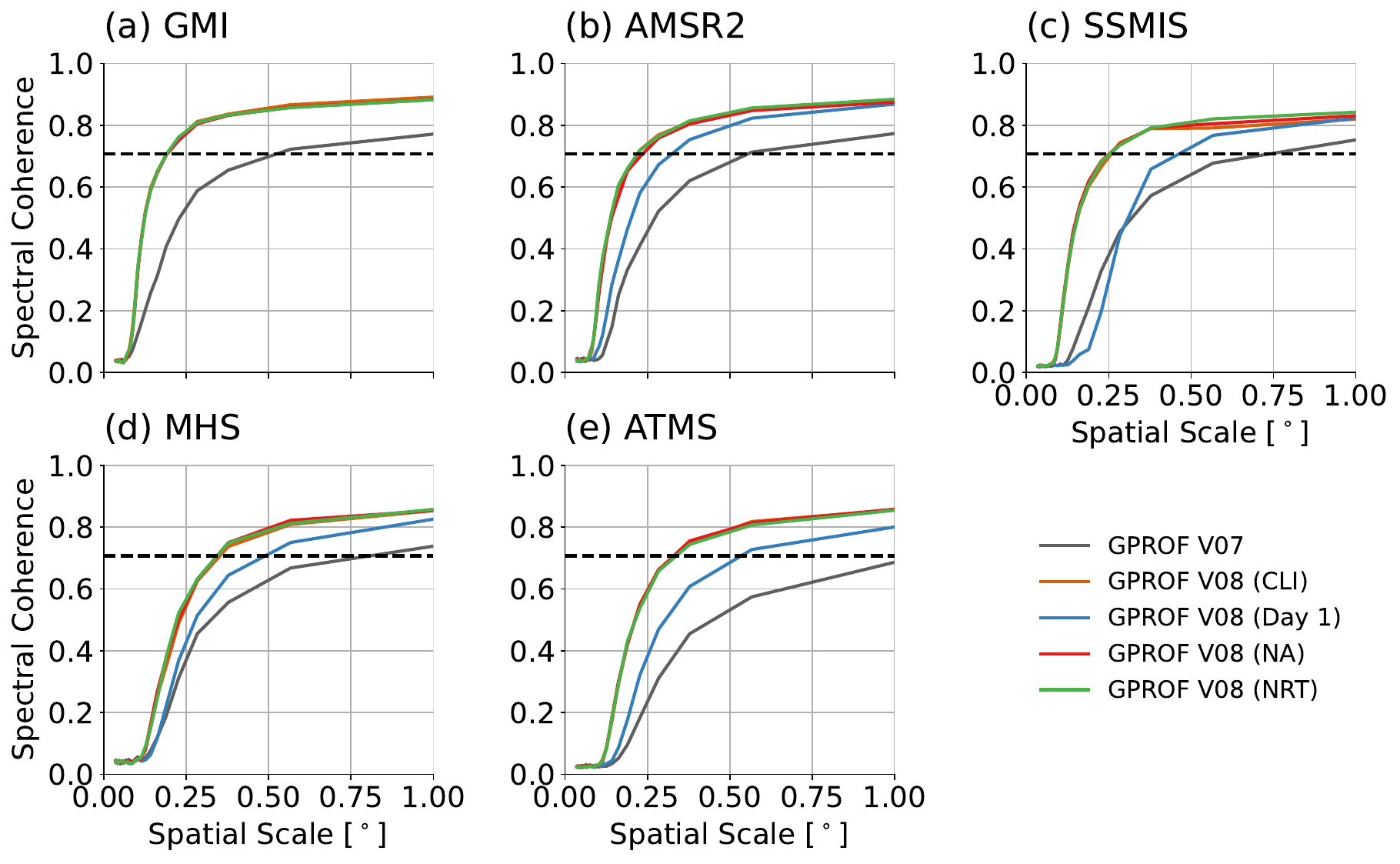}
    \caption{
      Spectral coherence curves for GPROF V07 and GPROF V08 retrievals compared to MRMS reference
      precipitation estimates using all overpasses from the year 2022. Panels show the spectral
      coherence curves for (a) GMI, (b) AMSR2, (c) SSMIS, (d) MHS, and (e) ATMS.
  }
  \label{fig:spectral_coherence}
\end{figure}

The continuous spatial coverage of the MRMS data also allows the effective
resolution of the retrieval to be analyzed. The effective resolution provides a
measure of the minimum size of precipitation features that the retrieval is able
to reproduce on average. The effective resolution differs from and is generally
coarser than the spatial sampling of the retrieval, which for GPROF corresponds to
the spatial sampling of the sensor footprints since retrievals are performed on
the radiometer swaths. \citet{pfreundschuh_gprof_2024} showed that the GPROF-NN
3D retrieval is able to more than double the effective resolution of GMI over
land. Fig.~\ref{fig:spectral_coherence} shows the spectral coherence of GMI,
AMSR2, SSMIS, MHS, and ATMS with respect to MRMS reference precipitation
measurements.

The results for GMI are similar to those reported in
\citet{pfreundschuh_gprof_2024} with GPROF V08 achieving an effective resolution
of around $0.19\unit{^\circ}$ -- more than double of the effective resolution of
$0.49^\circ$of GPROF V07. For MHS, the improvements are of similar order with
GPROF V08 improving the effective resolution to $0.35^\circ$ from $0.81^\circ$.
For both SSMIS and ATMS, the improvements are even  larger. For ATMS the
effective resolution is increased  to $0.33^\circ$ from $1.08^\circ$ for GPROF
V07 corresponding to an increase of a factor of three. For SSMIS, effective
resolution improves from $0.94$ to $0.24$ corresponding to an improvement of
more than a factor of $3.5$ compared to GPROF V07. While the Day-1 also improves the
effective resolution for all sensors, the relative improvements are only about
half of the improvements afforded by the finetuned retrievals. GPROF V08 is thus
able to achieve large improvements in the effective resolution of the retrieved
precipitation fields not only for GMI but also the other GPM constellation
sensors. Table~\ref{tab:effective_resolutions} lists the effective resolutions
for all assessed sensors and retrievals.

\begin{table}[hbpt]
  \caption{
    Effective resolution in $^\circ$ of the GPROF V08 and V07 retrievals for GMI, AMSR2, SSMIS on F17, and MHS on NOAA19
    with respected to MRMS calculated using all overpasses from 2022.
  }
  \label{tab:effective_resolutions}
  \begin{tabular}{l|ccccc}
    Sensor & GPROF V07 & GPROF V08 (Day 1) & GPROF V08 (NA) & GPROF V08 (NRT) & GPROF V08 (CLI) \\
    \hline
    GMI    & 0.49      &                   &  0.19          &  0.19           &  0.19           \\
    AMSR2  & 0.55      &  0.32             &  0.24          &  0.22           &  0.22           \\
    SSMIS  & 0.94      &  0.61             &  0.26          &  0.24           &  0.26           \\
    MHS    & 0.81      &  0.49             &  0.34          &  0.34           &  0.35           \\
    ATMS   & 1.08      &  0.53             &  0.33          &  0.34           &  0.33           \\
  \end{tabular}
\end{table}

\begin{figure}[hbpt]
  \centering
  \includegraphics[width=\textwidth]{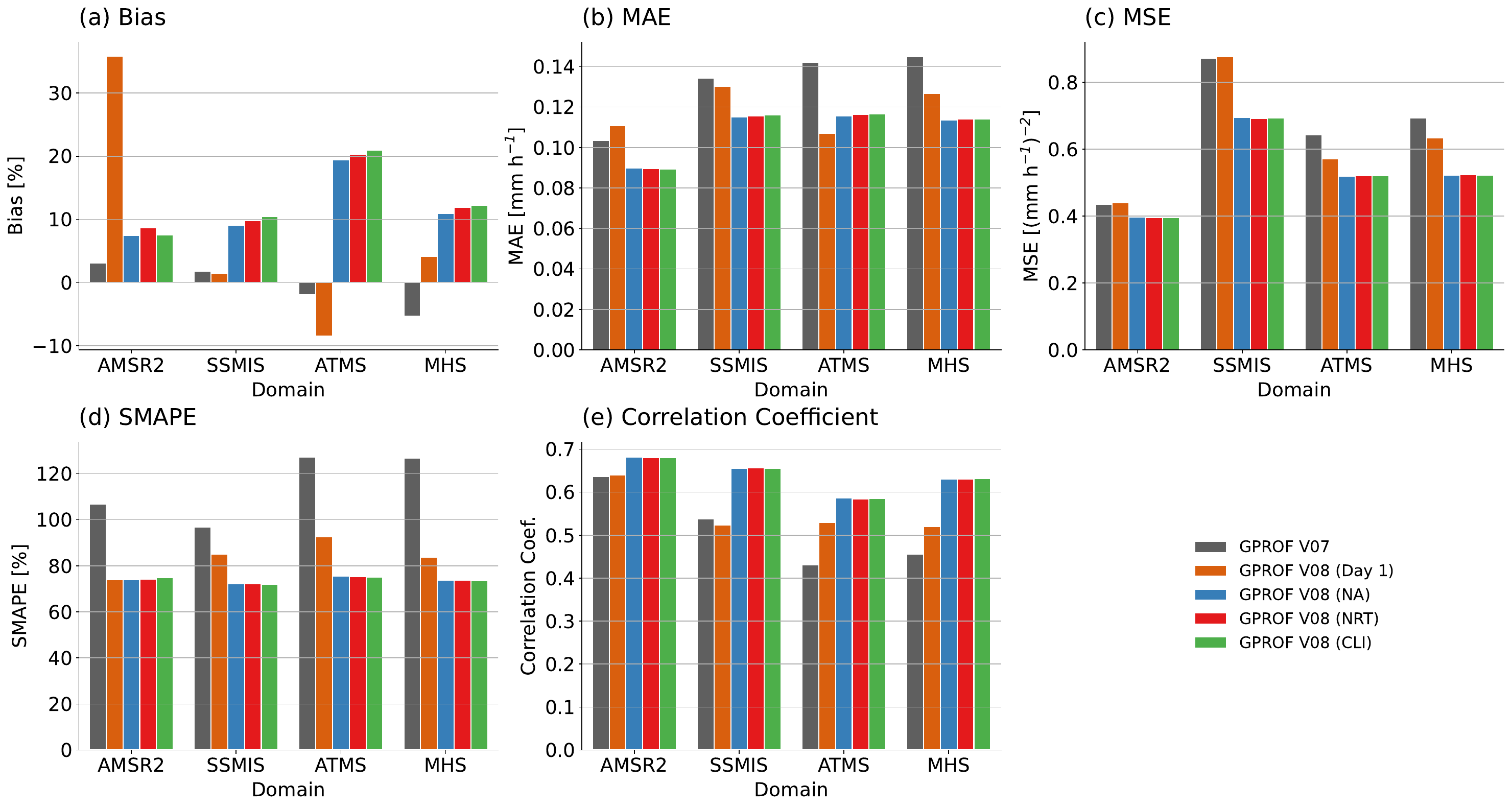}
    \caption{
      Retrieval accuracy of GPROF V08 for principal GPM-era constellation
      sensors with respect to  2BCMB precipitation estimates for the year
      2022.
  }
  \label{fig:validation_cmb}
\end{figure}

Corresponding results for the evaluation of GPM-era sensors against reference
precipitation measurements from the 2BCMB product are shown in
Fig.~\ref{fig:validation_cmb}. The GPROF V08 retrievals all exhibit positive
biases compared to 2BCMB. These biases are expected since the GPROF V08 training
uses the GPROF-NN XPR retrieval to add light precipitation that would be missed
by the DPR and thus the 2BCMB product. Although the GPROF V07 retrieval used
light precipitation estimates from MiRS to account for precipitation missed by
the DPR, this did not lead to a notable increase in overall precipitation with
its biases rangin from -5\% to 3\% across the four sensors.

In terms of the remaining error metrics, the finetuned GPROF V08 retrievals show
robust improvements over GPROF V07 for all sensors. The relative improvement are
relatively smallest for AMSR2 where they range from 6\% in linear correlation
coefficient to 30\% in SMAPE. For the two atmospheric sounders, MHS and ATMS,
the relative improvements are substantially larger ranging from 15\% for the MAE
to about 40\% for SMAPE and linear correlation coefficient. While the Day-1
retrieval yields improvements over GPROF V07 for the two cross-track scanners,
it remains at the level of GPROF V07 for the conical scanners, again
highlighting the importance of the finetuning step for exploiting the full
potential the observations for precipitation retrieval.

\subsubsection{TRMM-Era Sensors}

The validation of the GPROF V08 retrievals for TRMM-era sensors uses
precipitation estimates from the TRMM combined radar and radiometer retrievals
as reference precipitation estimates. The estimates are produced using the same
algorithm as the GPM 2BCMB and are therefore not fully independent of the GPROF
V08 training data. However, the combined TRMM radar and radiometer retrievals
are one of the few high-quality precipitation measurements available for
validating precipitation retrievals from early PMW sensors.

\begin{figure}[hbpt]
  \centering
  \includegraphics[width=\textwidth]{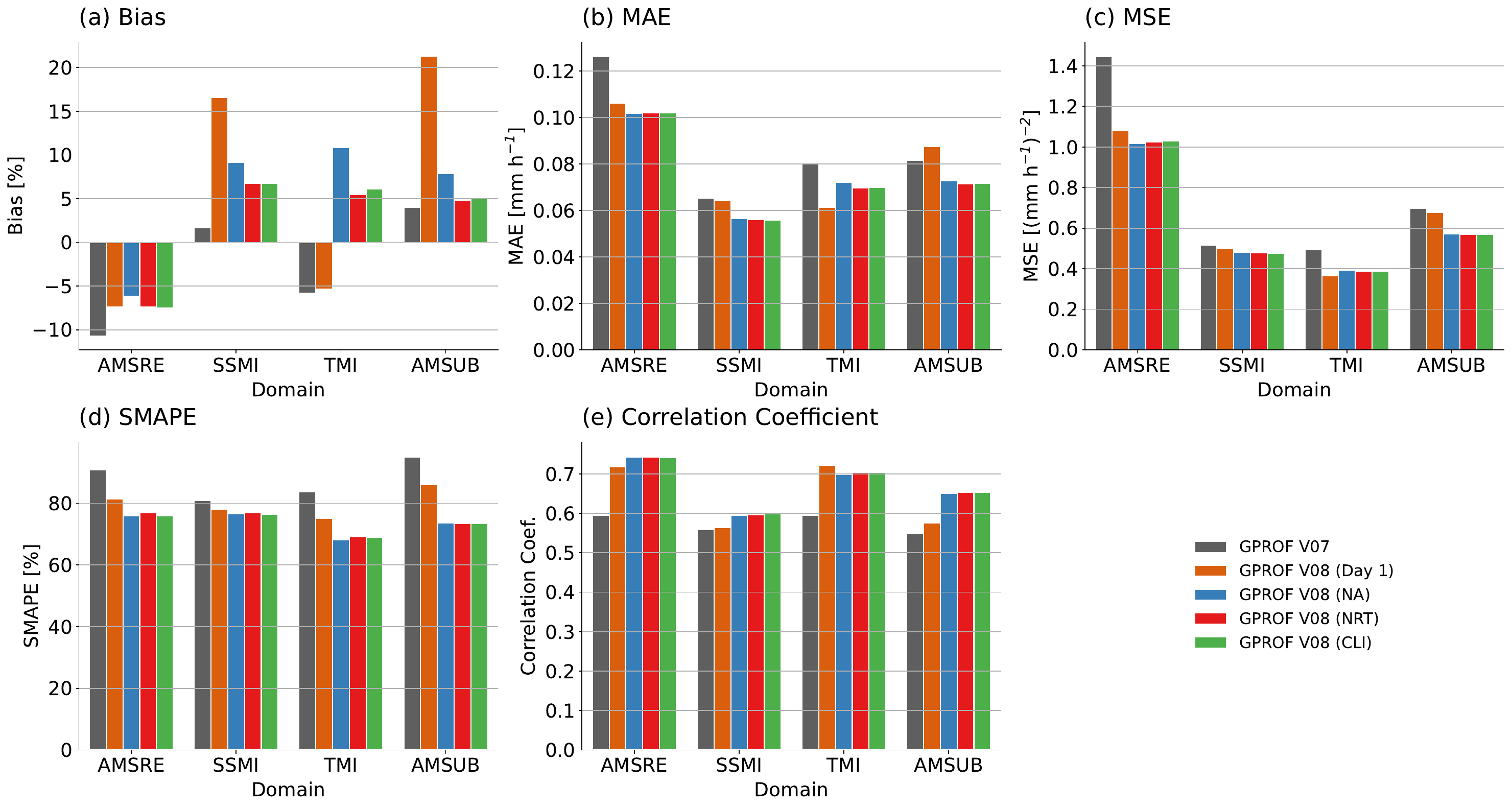}
    \caption{
    Retrieval accuracy of GPROF V08 for TRMM-era GPM constellation sensors with
    respect to combined radar and radiometer retrievals from the TRMM 2BCMB product
    over land.
  }
  \label{fig:validation_trmm_land}
\end{figure}

The accuracy of the GPROF V08 retrievals for AMSRE, SSMI, TMI, and AMSU-B over
land are shown in Fig.~\ref{fig:validation_trmm_land}. Retrieval biases remain
within 11\% for GPROF V07 and the finetuned GPROF V08 retrievals while the Day-1
retrieval exhibits larger biases for SSMI and AMSUB. In terms of retrieval
errors and linear correlation, the finetuned GPROF V08 retrievals again show
robust improvements over GPROF V07.

\begin{figure}[hbpt]
  \centering
  \includegraphics[width=\textwidth]{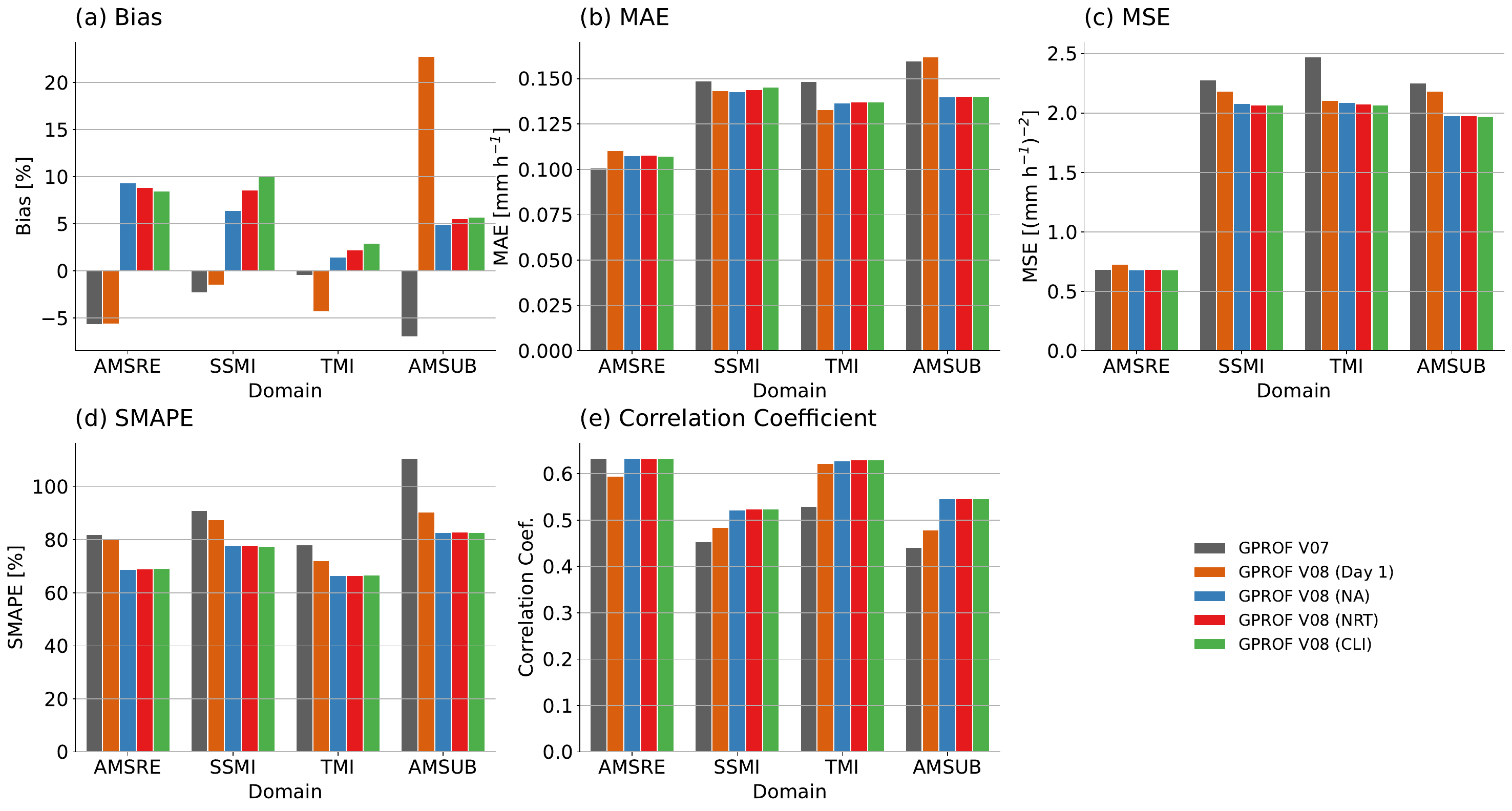}
    \caption{
Retrieval accuracy of GPROF V08 for TRMM-era GPM constellation sensors with
respect to combined radar and radiometer retrievals from the TRMM 2BCMB product
over ocean.
  }
  \label{fig:validation_trmm_ocean}
\end{figure}

The corresponding validation results over the ocean are shown in
Fig.~\ref{fig:validation_trmm_ocean}. All GPROF V08 retrievals exhibit a
tendency to overestimate the TRMM 2BCMB estimates albeit to a lower degree than
for the GPM-era sensors. This is likely due to the restriction of lower
latitudes imposed by the TRMM orbit. Also over ocean, the finetuned GPROF V08
retrievals reduce retrieval errors and increases the linear correlation with the
reference estimates compared to the GPROF V07 baselines.

\conclusions  

This two-part study describes the GPROF V08 retrieval -- the latest version of
the operational PMW precipitation retrievals for the GPM mission. GPROF V08
replaces the Bayesian retrieval scheme used in all prior versions of GPROF with
a ML retrieval based on a CNN. The principal advantage of the novel ML approach
is its ability to leverage spatial features in the input observations to
better constrain the observed precipitation fields. This allows the updated GPROF V08
retrieval to substantially improve retrieval accuracy across the full GPM
constellation. However, this change in retrieval paradigm brought with it novel
challenges for the operationalization of the retrieval.

The principal challenge of GPROF V08 was the generation of consistent training
data for the full constellation of current and historical GPM PMW radiometers.
CNN-based retrievals require training data consisting of two-dimensional image
patches. Because the conventional simulation framework used by GPROF was limited
to the central part of the DPR swath, its spatial coverage was too limited to
allow creating sufficiently large training scenes. To address this, we developed
a novel ML simulation framework for generating synthetic observations for any
GPM PMW sensor across the full GMI swath. The analysis presented in the first
part of this study demonstrated that the ML-based simulator provides more
accurate simulated observations than conventional radiative transfer
simulations, which has been shown crucial to avoid the mismatch between
simulated observations encountered during training and actual observations
encountered during inference from degrading the retrieval accuracy compared to
GPROF V07 retrievals. Nonetheless, GPROF-NN retrievals trained solely on
simulated observations were still found to be susceptible to climatological
biases of up to $20\ \unit{\%}$. Moreover, the retrievals trained on synthetic
observations were still found to lag behind retrievals trained on actual
observations in terms of accuracy. This shows that even the updated
simulation framework is not yet accurate enough to fully replace training on actual
observations.

The second part of this study presented the full operational ML training
pipeline of GPROF V08 and validated resulting surface precipitation estimates
against independent reference estimates from several ground-based datasets as
well as combined radar and radiometer retrievals from GPM and TRMM. The critical
innovation introduced in this second study is the finetuning using actual
observations. For non-GMI sensors, the GPROF V08 training pipeline produces two
retrievals: A Day-1 retrieval trained mainly on synthetic observations and the
final retrieval that has been finetuned using actual observations collocated
with precipitation estimates from a GPM reference sensor. The validation covered
all principal GPM sensors except MWI on WSFM-1 and AMSR3 on GoSat GW, which have
only come online during the last two years and have not yet produced
sufficiently many observations for a robust, independent validation outside the
finetuning period.

The comprehensive assessment of current and historical GPM sensors demonstrated
that improvements in retrieval accuracy achieved by GPROF V08 are robust across
all sensors and assessed metrics. The magnitude of the improvements varied
across the GPM sensors as well as surface types. The retrievals for AMSRE and
AMSR2 were already quite accurate in GPROF V07 (likely because of the similarity
to GMI) and the relative improvements therefore generally smaller, particularly
over ocean. For the atmospheric sounders (AMSUB, ATMS, and MHS) as well as
SSMIS, the improvements were substantially larger reaching up to 40 \% in the
linear correlation with 2BCMB reference estimates over ocean. This suggest that
the sensors that benefit most from the updated ML-based retrievals are those
that were traditionally more difficult to simulate because of large footprints
or cross-track-scanning viewing geometries. The analysis of the effective
resolution of the retrievals  over CONUS furthermore
showed that GPROF V08 increases the effective resolution of the PMW
precipitation fields by a factor of two for GMI, AMSR2, and MHS, and a factor of
three or more for SSMIS and ATMS.

The analysis further showed that the ancillary data has no significant impact on
the retrieval accuracy. The benefit of ancillary remained negligible even for
the atmospheric sounders of the constellation despite their observations
providing generally lower information content than dedicated precipitation
sensors such as GMI or AMSR2. The reliance on ancillary data not only increases
the data dependencies of the retrieval but the extraction and preparation also
takes a non-negligible amount of the total retrieval processing time. Our
results show that the GPROF V08 retrieval can be run directly on GPM L1C
observations without notably degrading the accuracy. Since we make both the
GPROF V08 retrieval implementation and trained model weights openly available,
interested users can directly run the GPROF V08 retrieval on the publicly
available GPM L1C files. GPROF V08 is therefore the first GPROF retrieval that
is truly, openly accessible to the scientific community.

By leveraging AI-driven retrieval techniques, GPROF V08 is able to improve
global precipitation estimates consistently across the full GPM PMW-sensor
constellation. Because of the important role the these observations play for
global precipitation monitoring, GPROF V08 constitutes a an important step
towards improved global precipitation estimates.


\codedataavailability{
  The implementation of the GPROF V08 retrieval and the model weights are available through
  the \texttt{gprof\_nn} Python package \citep{gprof_nn_github}.
}

\noappendix       




\appendixfigures  

\appendixtables   


\authorcontribution{
    SP has developed the GPROF-NN retrieval framework. SP, PB, and CK have
developed the operational GPROF V08 retrieval. SP has performed the validation
of the retrieval results and compiled manuscript. PB and CK provided feedback on
several iteration of the manuscript draft.

} 

\competinginterests{No competing interests are present.} 

\begin{acknowledgements}
The work of SP, PB, and CK has been supported by NASA grant
80NSSC22K0604.
\end{acknowledgements}



\bibliographystyle{copernicus}
\bibliography{references.bib}

\end{document}